\documentclass[12pt]{article}
\usepackage{graphicx}
\usepackage{latexsym}
\usepackage{amscd}
\usepackage[cp1251]{inputenc}
\usepackage[english,russian]{babel}
\usepackage{amsmath,  eucal}
\usepackage{amssymb}
\usepackage{indentfirst}
\usepackage[small,centerlast]{caption2}
\usepackage{longtable}
\usepackage[dvipsnames]{color}
\usepackage{cite}
\graphicspath{{figure/}}

\makeatletter
\newcommand{\const}{\mathop{\rm const\, }}
\renewcommand{\Re}{\mathop{\rm Re\,}}

\makeatother {\renewcommand{\baselinestretch}{1.1}

\makeatletter
\renewcommand{\section}{\@startsection {section}{1}{\z@}%
                                   {-3.5ex \@plus -1ex \@minus -.2ex}%
                                   {2.3ex \@plus.2ex}%
                                   {\normalfont\Large\uppercase}}
\renewcommand{\subsection}{\@startsection{subsection}{2}{\z@}%
                                     {-3.25ex\@plus -1ex \@minus -.2ex}%
                                     {1.5ex \@plus .2ex}%
                                     {\normalfont\large\itshape}}
\renewcommand{\subsubsection}{\@startsection{subsubsection}{3}{1em}%
                                     {-3.25ex\@plus -1ex \@minus -.2ex}%
                                     {-1.5em \@plus .2em}%
                                     {\normalfont\normalsize\bfseries}}

\makeatother

\begin{document}

\newcommand{\mc}[1]{\mathcal{#1}}
\newcommand{\E}{\mc{E}}
\thispagestyle{empty} \large
\renewcommand{\abstractname}{}

 \begin{center}
{\Large \textbf{Longitudinal and transversal current in collisional
plasma, generated by two transversal electromagnetic waves}}
\medskip

{\large \textbf{A. V. Latyshev, A. A. Yushkanov}}
\medskip

\textit{Moscow State Regional University}
\\[10pt]

\end{center}

\begin{abstract}
\noindent\small
From kinetic Vlasov equation for collisional plasmas distribution
function  in square-law approximation on sizes of intensivities
of electric fields is received. The known integral of collisions
of relaxation type, so-called BGK (Bhatnagar, Gross, Krook)
integral of collisions is considered.
The formula for calculation electric current
at any temperature (any degree of degeneration of
electronic gas) is deduced.
This formula contains an one-dimensional quadrature.

It is shown, that the nonlinearity account leads to occurrence
the longitudinal electric current directed along a wave vector.
This longitudinal current is orthogonal to a known transversal
classical current, received at the linear analysis.

When frequency of collisions tends  to the zero, all received
results for collisional plasmas pass in corresponding formulas for
collisionless plasmas.

The case of small values of wave number is considered.
It is shown, that the received quantity of  longitudinal current
at aspiration of frequency of collisions to zero also passes
in corresponding expression of current for collisionless plasmas.
Graphic comparison of dimensionless quantity of  current
depending on wave number, frequency of oscillations of electromagnetic field
and frequencies of electron collisions with plasma particles
is carry out.
\end{abstract}

\section*{ВВЕДЕНИЕ}

В настоящей работе выводятся формулы для вычисления электрического
тока в столкновительной плазме Ферми---Дирака.

При решении кинетического уравнения Власова, описывающего
поведение плазмы, мы учитываем как в разложении
функции распределения, так и в разложении  величины самосогласованного
электромагнитного поля величины, пропорциональные
квадратам напряженностей внешних электрических полей и их произведению.

При таком нелинейном подходе оказывается, что электрический ток
имеет две ненулевые компоненты. Одна компонента электрического
тока направлена вдоль напряженности электрического поля.
Эта компонента электрического поля
точно такая же, как и в линейном анализе. Это "поперечный"\,
ток.
Таким образом, в линейном приближении мы получаем известное
выражение поперечного электрического тока.

Вторая ненулевая компонента электрического тока имеет второй
порядок малости относительно величин напряженностей
электрических полей.
Вторая компонента электрического тока направлена вдоль волнового
вектора. Этот ток
ортогонален первой компоненте. Это "продольный"\, ток.

Появление продольного тока выявляется проведенным
нелинейным анализом взаимодействия электромагнитных полей
с плазмой.

Нелинейные эффекты в  плазме изучаются уже длительное время
\cite{Gins}--\cite{Lat1}.

В работах \cite{Gins} и \cite{Zyt} изучаются нелинейные эффекты в
плазме. В работе \cite{Zyt} нелинейный ток использовался, в
частности, в вопросах вероятности распадных процессов. Отметим,
что в работе \cite{Zyt2} указывается на существование
нелинейного тока вдоль волнового вектора (см. формулу (2.9) из \cite{Zyt2}).

Квантовая плазма изучалась в работах \cite{Lat1}--\cite{Lat8}.
Столкновительная квантовая плазма начала изучаться в работе
Мермина \cite{Mermin}. Затем квантовая столкновительная плазма
изучалась в наших работах \cite{Lat2}--\cite{Lat6}. Нами
изучалась квантовая столкновительная плазма с переменной частотой
столкновений. В работах    \cite{Lat7} -- \cite{Lat9} было
исследовано генерирование продольного тока поперечным
электромагнитным полем в классической и квантовой плазме
Ферми---Дирака \cite{Lat7}, в максвелловской плазме
\cite{Lat8} и в вырожденной плазме \cite{Lat9}.

Укажем еще на ряд работ по  плазме, в том числе и  квантовой.
Это работы \cite{Andres}--\cite{Manf}.

В настоящей работе рассматривается классическая плазма
при произвольной температуре, то-есть при произвольной
степени вырождения электронного газа.
В плазме распространяются два внешних электромагнитных поля.
Требуется найти отклик плазмы на эти два поля. Для этого
решается уравнение Власова---Больцмана методом малого параметра.
В качестве интеграла столкновений используется модель
релаксационного типа, так называемый интеграл столкновений БГК
(Бхатнагар, Гросс и Крук)
(см., например, \cite{BGK} и \cite{Welander}).
В качестве малых параметров рассматриваются величины
напряженностей электрических полей. Построена функция
распределения электронов плазмы. Для этого используется
квадратичное (по степени малых параметров) разложение функции
распределения и интеграла столкновений.

Функция распределения используется для для вычисления
электрического тока в плазме. Оказалось, что генерируемый двумя
электромагнитными полями электрический ток в плазме имеет,
как уже указывалось, две
ненулевые ортогональные компоненты. Одна компонента --- это
известный поперечный ток. Этот ток выявляется при линейном
анализе с помощью линейного разложения функции распределения и
интеграла столкновений. Квадратичное разложение функции
распределения и интеграла столкновений не влияет на величину и
направление линейного поперечного тока. Помимо линейного тока
нелинейный анализ обнаруживает существование продольного
электрического тока, ортогонального известному классическому току.
Этот продольный ток состоит из трех  и определяется квадратами
напряженностей электрических полей и произведением их первых
степеней. Первые два слагаемых продольного электрического тока
(пропорциональные квадратам напряженностей электрических полей)
имеют такую же структуру, что  и в случае одного электрического
тока (см. \cite{Lat7}). Принципиально новым составляющим током
является так называемый "перекрестный"\, ток. Это слагаемое,
пропорциональное произведению напряженностей электрических
полей.

В заключение статьи приводится графический анализ
"перекрестного"\, тока: приведены графики действительной и
мнимой части безразмерного перекрестного тока в зависимости от
величины волнового числа одной из электрических волн при
различных значениях химического потенциала, частоты столкновений
и волнового числа второй из электрических волн.

\section{Уравнение Власова---Больцмана}

Покажем, что в случае классической плазмы, описываемой уравнением
Власова, генерируется продольный ток и вычислим его плотность. На
существование этого тока указывалось более полувека тому назад
\cite{Zyt2}.

Возьмем уравнение Власова, описывающее поведение
столкновительной плазмы с интегралом столкновений БГК
(Бхатнагар, Гросс и Крук)
$$
\dfrac{\partial f}{\partial t}+\mathbf{v}\dfrac{\partial f}{\partial
\mathbf{r}}+
e\bigg(\mathbf{E}_1+{\mathbf E}_2+
\dfrac{1}{c}[\mathbf{v},\mathbf{H}_1+{\mathbf H}_2]\bigg)
\dfrac{\partial f}{\partial\mathbf{p}}=\nu(f_{eq}-f).
\eqno{(1.1)}
$$

В уравнении (1.1) $f$ -- функция распределения электронов
плазмы, ${\bf E}_j, {\bf H}_j\quad (j=1,2)$ -- компоненты
электромагнитного поля,
$c$ -- скорость света, ${\bf p}=m{\bf v}$ -- импульс электронов,
${\bf v}$ -- скорость электронов, $\nu$ -- эффективная
частота столкновений электронов с частицами плазмы,
функция распределения электронов Ферми---Дирака,
$f_{eq}({\bf r},v)$ (eq$\equiv$ equi\-lib\-rium)
-- локально равновесное распределение Ферми---Дирака,
$$
f_{eq}({\bf r},v)=\Big[1+\exp\dfrac{\E-\mu({\bf
r})}{k_BT}\Big]^{-1}.
$$

Здесь
$\E={mv^2}/{2}$ -- энергия электронов, $\mu({\bf r})$ -- химический
потенциал электронного газа, $k_B$ -- постоянная Больцмана, $T$
-- температура плазмы.

Локально равновесное распределение Ферми---Дирака будем
использовать с безразмерными параметрами в виде
$$
f_{eq}({\bf r},P)=\big[1+\exp(P^2-\alpha({\bf r}))\big]^{-1}.
$$

Здесь
${\bf P}={{\bf P}}/{p_T}$ --
безразмерный импульс электронов,  $p_T=mv_T$,
$v_T$ -- тепловая скорость электронов,
$\alpha({\bf r})$ -- безразмерный химический потенциал,
$k_BT=\E_T$ -- тепловая кинетическая энергия электронов,

$$
v_T=\sqrt{\dfrac{2k_BT}{m}},\qquad
\alpha({\bf r})=\dfrac{\mu({\bf r})}{k_BT}, \qquad
k_BT=\E_T=\dfrac{mv_T^2}{2}.
$$

Ниже нам понадобится абсолютное распределение Ферми---Дирака
$f_0(v)$ с постоянным химическим потенциалом,
$$
f_0(v)=\Big[1+\exp\dfrac{\E-\mu}{k_BT}\Big]^{-1}=
\big[1+\exp(P^2-\alpha)\big]^{-1}=f_0(P),
$$
где $\alpha=\const$.

Будем считать, что в плазме имеется два электромагнитных поля,
каждое из которых представляет собой бегущую гармоническую волну
с волновым вектором ${\bf k}_j$ и частотой колебаний $\omega_j$:
$$
{\bf E}_j={\bf E}_{0j}e^{i({\bf k_jr}-\omega_j t)}, \qquad
{\bf H}_j={\bf H}_{0j}e^{i({\bf k_jr}-\omega_j t)}, \quad j=1,2.
$$

Напряженности электрических и магнитных полей связаны между собой через
соответствующие векторные потенциалы электромагнитных полей:
$$
\mathbf{E}_j=-\dfrac{1}{c}\dfrac{\partial \mathbf{A}_{j}}{\partial
t}=\dfrac{i\omega_j}{c}\mathbf{A}_j,\;\qquad
\mathbf{H}_j={\rm rot} \mathbf{A}_j,\quad j=1,2.
$$

Будем считать, что векторный потенциал электромагнитного поля
${\bf A}_j({\bf r},t)$ ортогонален волновому вектору ${\bf
k}_j$, т.е.
$$
{\bf k}_j\cdot {\bf A}_j({\bf r},t)=0,\qquad j=1,2.
$$
Это значит, что волновой вектор ${\bf k}_j$ ортогонален
напряженностям соответствующих электрических и магнитных полей:
$$
{\bf k}_j\cdot {\bf E}_j({\bf r},t)=
{\bf k}_j\cdot {\bf H}_j({\bf r},t)=0,\qquad j=1,2.
$$

Для определенности будем считать, что волновые векторы обеих
полей направлены вдоль оси $x$, а электромагнитные поля
направлены вдоль оси $y$,
т.е.
$$
{\bf k}_j=k_j(1,0,0), \qquad {\bf E}_j=E_{j}(x,t)(0,1,0).
$$

Следовательно,
$$
{\bf E}_j=-\dfrac{1}{c}\dfrac{\partial {\bf  A}_j}{\partial t}=
\dfrac{i\omega_j}{c}{\bf A}_j,
$$
$$
{\bf H}_j=\dfrac{ck_j}{\omega_j}E_{j}\cdot(0,0,1),\qquad
{\bf [v,H}_j]=\dfrac{ck_j}{\omega_j}E_{j}\cdot (v_y,-v_x,0),
$$\medskip
$$
E_j=E_j(x,t)=E_{0j}e^{i(k_jx-\omega_j t)}\quad (j=1,2),
$$ \medskip
$$
e\bigg(\mathbf{E}_j+\dfrac{1}{c}[\mathbf{v},\mathbf{H}_j]\bigg)
\dfrac{\partial f}{\partial\mathbf{p}}=
e\dfrac{E_{j}}{\omega_j}\Big[k_jv_y\dfrac{\partial f}{\partial p_x}+
(\omega_j-k_jv_x)\dfrac{\partial f}{\partial p_y}\Big],
$$ \medskip
а также
$$
[{\bf v, H}_j]\dfrac{\partial f_0}{\partial {\bf p}}=0, \quad
\text{так как}\quad \dfrac{\partial f_0}{\partial {\bf p}}\sim
{\bf v}.
$$

Теперь уравнение (1.1) можно представить в следующем виде:
$$
\dfrac{\partial f}{\partial t}+v_x\dfrac{\partial f}{\partial x}+
\nu[f-f_{eq}]=-e\sum\limits_{j=1}^{2}\dfrac{E_{j}}{\omega_j}
\Big[k_jv_y\dfrac{\partial f}{\partial p_x}+
(\omega_j-k_jv_x)\dfrac{\partial f}{\partial p_y}\Big]
\eqno{(1.2)}
$$

Рассмотрим линеаризацию локально равновесной функции
распределения
$$
f_{eq}(P,x)=f_0(P)+g(P)\delta \alpha(x),
$$
где
$$
f_0(P)=\big[1+e^{P^2-\alpha}\big]^{-1},
$$
$$
\alpha(x)=\alpha+\delta \alpha(x),\qquad \alpha=\const,
$$
$$
g(P)=\dfrac{\partial f_0(P)}{\partial \alpha}=
\dfrac{e^{P^2-\alpha}}{(1+e^{P^2-\alpha})^2}.
$$

Уравнение (1.1) может быть переписано в виде
$$
\dfrac{\partial f}{\partial t}+v_x\dfrac{\partial f}{\partial x}+
\nu[f-f_0]=-
e\sum\limits_{j=1}^{2}\dfrac{E_j}{\omega_j}
\Big[k_jv_y\dfrac{\partial f}{\partial p_x}+
(\omega_j-k_jv_x)\dfrac{\partial f}{\partial p_y}\Big]+
$$
$$
+\nu g(P)\delta \alpha(x).
\eqno{(1.3)}
$$

Величина $ \delta \alpha (x) $ может быть найдена из закона
сохранения числа частиц
$$
\int (f_{eq}-f)\dfrac{2d^3p}{(2\pi \hbar)^2}=0.
$$

Из этого закона сохранения мы получаем
$$
\delta \alpha(x)\int g(P)\dfrac{2d^3p}{(2\pi \hbar)^2}=\int
[f-f_0(P)]
\dfrac{2d^3p}{(2\pi \hbar)^2}.
$$
Отсюда находим, что
$$
\delta \alpha(x)=\dfrac{\displaystyle\int [f-f_0(P)] d^3P}
{\displaystyle\int g(P)d^3P}.
$$

Заметим, что
$$
\int g(P)d^3P=2\pi \int\limits_{0}^{\infty}\dfrac{dP}{1+e^{P^2-\alpha}}=
\pi \int\limits_{-\infty}^{\infty}\dfrac{dP}{1+e^{P^2-\alpha}}=
\pi \hat f_0(\alpha),
$$
где
$$
\hat f_0(\alpha)= \int\limits_{-\infty}^{\infty}\dfrac{dP}{1+e^{P^2-\alpha}}
=2\int\limits_{0}^{\infty}\dfrac{dP}{1+e^{P^2-\alpha}}.
$$

Следовательно
$$
\delta \alpha(x)=\dfrac{1}{\pi\hat f_0(\alpha)}\int [f-f_0(P)] d^3P.
$$

Уравнение $(1.3)$ может быть преобразовано теперь к интегральному уравнению
$$
\dfrac{\partial f}{\partial t}+v_x\dfrac{\partial f}{\partial x}+
\nu [f-f_0(P)]=-e\sum\limits_{j=1}^{2}\dfrac{E_j}{\omega_j}
\Big[k_jv_y\dfrac{\partial f}{\partial p_x}+
(\omega_j-k_jv_x)\dfrac{\partial f}{\partial p_y}\Big]+
$$
$$
+\nu g(P)\dfrac{1}{\pi\hat f_0(\alpha)}\int [f-f_0(P)] d^3P.
\eqno{(1.4)}
$$

\section{Первое приближение}

Теперь действуем методом последовательных приближений, считая
малыми параметрами величины напряженностей электрических полей
$E_1$ и $E_2$.
Будем искать решение уравнения (1.4) в виде
$$
f=f_0(P)+f_1+f_2,
\eqno{(2.1)}
$$
где $f_1$ -- линейная комбинация первых степеней электрических
волн, а $f_2$ -- величина первого порядка малости относительно
$f_1$.

Теперь уравнение (1.4) с помощью (2.1) эквивалентно следующим уравнениям
$$
\dfrac{\partial f_1}{\partial t}+
v_x\dfrac{\partial f_1}{\partial x}+\nu f_1=
$$
$$
=-e\sum\limits_{j=1}^{2}\dfrac{E_j}{\omega_j}
\Big[k_jv_y\dfrac{\partial f_0}{\partial p_x}+
(\omega_j-k_jv_x)\dfrac{\partial f_0}{\partial p_y}\Big]
+\nu g(P)\dfrac{1}{\pi \hat f_0(\alpha)}\int f_1 d^3P.
\eqno{(2.2)}
$$ \bigskip
и
$$
\dfrac{\partial f_2}{\partial t}+
v_x\dfrac{\partial f_2}{\partial x}+\nu f_2=
$$
$$
=-e\sum\limits_{j=1}^{2}\dfrac{E_j}{\omega_j}
\Big[k_jv_y\dfrac{\partial f_1}{\partial p_x}+
(\omega_j-k_jv_x)\dfrac{\partial f_1}{\partial p_y}\Big]
+\nu g(P)\dfrac{1}{\pi \hat f_0(\alpha)}\int f_2 d^3P.
\eqno{(2.3)}
$$ \bigskip

В качестве $f_1$ возьмем
$$
f_1=E_1\varphi_1+E_2\varphi_2,
\eqno{(2.4)}
$$

Из уравнения (2.2) с помощью (2.4) получаем два уравнения
$$
[\nu-i\omega_j+ik_jv_x]\varphi_j=
$$
$$
=-e\dfrac{1}{\omega_j}
\Big[k_jv_y\dfrac{\partial f_0}{\partial p_x}+
(\omega_j-k_jv_x)\dfrac{\partial f_0}{\partial p_y}\Big]
+\nu g(P)A_j, \qquad j=1,2.
\eqno{(2.5)}
$$

Здесь
$$
A_j=\dfrac{1}{\pi \hat f_0(\alpha)}\int \varphi_j d^3P,\qquad j=1,2.
\eqno{(2.6)}
$$

Введем безразмерные параметры
$$
\Omega_j=\dfrac{\omega_j}{k_Tv_T},\qquad y=\dfrac{\nu}{k_Tv_T},
\qquad q_j=\dfrac{k_j}{k_T}.
$$

Здесь $q_j$ -- безразмерное волновое число,
$k_T ={mv_T}/{\hbar} $ -- тепловое волновое число, $ \Omega_j$
-- безразмерная частота колебаний электромагнитного поля ${\bf
E}_j$, $y$ -- безразмерная частота столкновений электронов с
частицами плазмы.

В уравнении (2.5) перейдем к безразмерным параметрам. В
результате получаем уравнение
$$
i(q_jP_x-z_j)\varphi_j=$$$$=-\dfrac{e}{k_Tp_Tv_T\Omega_j}
\Big[q_jP_y\dfrac{\partial f_0}{\partial P_x}+
(\Omega_j-q_jP_x)\dfrac{\partial f_0}{\partial P_y}\Big]
+y g(P)A_j.
\eqno{(2.7)}
$$

Здесь
$$
z_j=\Omega_j+iy=\dfrac{\omega_j+i \nu}{k_Tv_T},\qquad j=1,2.
$$

Заметим, что
$$
\dfrac{\partial f_0}{\partial P_x}\sim P_x,\qquad
\dfrac{\partial f_0}{\partial P_y}\sim P_y.
$$

Следовательно
$$
\Bigg[q_jP_y\dfrac{\partial f_0}{\partial P_x}+
(\Omega_j-q_jP_x)\dfrac{\partial f_0}{\partial P_y}\Bigg]=
\Omega_j\dfrac{\partial f_0}{\partial P_y}.
$$

Теперь из уравнения (2.7) находим, что
$$
\varphi_1=\dfrac{ie}{k_Tp_Tv_T}\cdot\dfrac{\partial f_0/\partial P_y}
{q_jP_x-z_j}-iy\cdot\dfrac{g(P)}{q_jP_x-z_j}A_j.
\eqno{(2.8)}
$$

Подставляя (2.8) в уравнение (2.6), получаем равенство
$$
A_1\Bigg(\pi\hat f_0(\alpha)+iy\int \dfrac{g(P)d^3P}
{q_jP_x-z_j}\Bigg)=\dfrac{ie}{k_Tp_Tv_T}
\int\dfrac{\partial f_0/\partial P_y}{qP_x-z}d^3P.
$$

Легко видеть, что интеграл в правой части этого равенства равен
нулю. Следовательно,
$$A_1=0.$$

Таким образом, согласно (2.8) функция $f_1$  построена
и определяется равенством
$$
f_1=\dfrac{ie}{k_Tp_Tv_T}\cdot\Bigg[\dfrac{E_1}{q_1P_x-z_1}+
\dfrac{E_2}{q_2P_x-z_2}\Bigg]\dfrac{\partial f_0}{\partial P_y}.
\eqno{(2.9)}
$$

\section{Второе приближение}

В правой части уравнения (2.3) перейдем к безразмерным
параметрам:
$$
\dfrac{\partial f_2}{\partial t}+
v_x\dfrac{\partial f_2}{\partial x}+\nu f_2-
\nu g(P)\dfrac{1}{\pi \hat f_0(\alpha)}\int f_2 d^3P=
$$
$$
=-\dfrac{e}{p_T}\sum\limits_{j=1}^{2}\dfrac{E_j}{\Omega_j}
\Big[q_jP_y\dfrac{\partial f_1}{\partial P_x}+
(\Omega_j-q_jP_x)\dfrac{\partial f_1}{\partial P_y}\Big].
\eqno{(3.1)}
$$ \bigskip

Вместо $f_1$ в уравнение (3.1) подставим (2.9). Получим уравнение
$$
\dfrac{\partial f_2}{\partial t}+
v_x\dfrac{\partial f_2}{\partial x}+\nu f_2-
\nu g(P)\dfrac{1}{\pi \hat f_0(\alpha)}\int f_2 d^3P=
$$
$$
=-\dfrac{ie^2}{k_Tp_T^2v_T}\Bigg\{\dfrac{E_1^2}{\Omega_1}\Bigg[
q_1P_y\dfrac{\partial}{\partial P_x}
\Big(\dfrac{\partial f_0/\partial P_y}{q_1P_x-z_1}\Big)+
\dfrac{\Omega_1-q_1P_x}{q_1P_x-z_1}\cdot\dfrac{\partial^2 f_0}{\partial P_y^2}
\Bigg]+
$$
$$
+\dfrac{E_1E_2}{\Omega_1}\Bigg[
q_1P_y\dfrac{\partial}{\partial P_x}
\Big(\dfrac{\partial f_0/\partial P_y}{q_2P_x-z_2}\Big)+
\dfrac{\Omega_1-q_1P_x}{q_2P_x-z_2}\cdot\dfrac{\partial^2 f_0}{\partial P_y^2}
\Bigg]+
$$
$$
+\dfrac{E_1E_2}{\Omega_2}\Bigg[q_2P_y\dfrac{\partial}{\partial P_x}
\Big(\dfrac{\partial f_0/\partial P_y}{q_1P_x-z_1}\Big)+
\dfrac{\Omega_2-q_2P_x}{q_1P_x-z_1}\cdot\dfrac{\partial^2 f_0}{\partial P_y^2}
\Bigg]+
$$
$$
+\dfrac{E_2^2}{\Omega_2}\Bigg[
q_2P_y\dfrac{\partial}{\partial P_x}
\Big(\dfrac{\partial f_0/\partial P_y}{q_2P_x-z_2}\Big)+
\dfrac{\Omega_2-q_2P_x}{q_2P_x-z_2}\cdot\dfrac{\partial^2 f_0}{\partial P_y^2}
\Bigg]\Bigg\}.
\eqno{(3.2)}
$$

В левую часть уравнения (3.2) подставим вместо $f_2$ выражение
$$
f_2=E_1^2\psi_1+E_2^2\psi_2+E_1E_2\psi_0.
\eqno{(3.3)}
$$

Получим три уравнения
$$
(\nu-2i\omega_j+2ik_jv_x)\psi_j-
\nu g(P)\dfrac{1}{\pi \hat f_0(\alpha)}\int \psi_j d^3P=
$$
$$
=-\dfrac{ie^2}{k_Tp_T^2v_T}\cdot\dfrac{1}{\Omega_j}\Bigg[
q_jP_y\dfrac{\partial}{\partial P_x}
\Big(\dfrac{\partial f_0/\partial P_y}{q_jP_x-z_j}\Big)+
\dfrac{\Omega_j-q_jP_x}{q_jP_x-z_j}\cdot\dfrac{\partial^2 f_0}{\partial P_y^2}
\Bigg]
\eqno{(3.4)}
$$
и
$$
[\nu-i(\omega_1+\omega_2)+i(k_1+k_2)v_x]\psi_0-
\nu g(P)\dfrac{1}{\pi \hat f_0(\alpha)}\int \psi_0 d^3P=
$$
$$
=-\dfrac{ie^2}{k_Tp_T^2v_T}\Bigg\{\dfrac{1}{\Omega_1}\Bigg[
q_1P_y\dfrac{\partial}{\partial P_x}
\Big(\dfrac{\partial f_0/\partial P_y}{q_2P_x-z_2}\Big)+
\dfrac{\Omega_1-q_1P_x}{q_2P_x-z_2}\cdot\dfrac{\partial^2 f_0}{\partial P_y^2}
\Bigg]+
$$
$$
+\dfrac{1}{\Omega_2}\Bigg[q_2P_y\dfrac{\partial}{\partial P_x}
\Big(\dfrac{\partial f_0/\partial P_y}{q_1P_x-z_1}\Big)+
\dfrac{\Omega_2-q_2P_x}{q_1P_x-z_1}\cdot\dfrac{\partial^2 f_0}{\partial P_y^2}
\Bigg]\Bigg\}.
\eqno{(3.5)}
$$

Из уравнения (3.4) находим:
$$
\psi_j=-\dfrac{iy}{2}\cdot \dfrac{g(P)}{q_jP_x-z_j'}B_j-
$$
$$
-\dfrac{e^2}{2k_T^2p_T^2v_T^2}\cdot \dfrac{1}{\Omega_j}
\Bigg[q_jP_y\dfrac{\partial}{\partial P_x}
\Big(\dfrac{\partial f_0/\partial P_y}{q_jP_x-z_j}\Big)+
\dfrac{\Omega_j-q_jP_x}{q_jP_x-z_j}\cdot
\dfrac{\partial^2 f_0}{\partial P_y^2}\Bigg]\dfrac{1}{q_jP_x-z_j'},
\eqno{(3.6)}
$$
где
$$
z_j'=\Omega_j+\dfrac{iy}{2}=\dfrac{\omega_j}{k_Tv_T}+i\dfrac{\nu}{2k_Tv_T}=
\dfrac{\omega_j+i \nu/2}{k_Tv_T},
$$\medskip
$$
B_j=\dfrac{1}{\pi \hat f_0(\alpha)}\int \psi_j d^3P,\qquad
j=1,2.
\eqno{(3.7)}
$$

Из уравнения (3.5) находим:
$$
\psi_0=-\dfrac{iy}{2}\cdot \dfrac{g(P)}{qP_x-z'}B_0-
$$
$$
-\dfrac{e^2}{2k_T^2p_T^2v_T^2}\Bigg\{\dfrac{1}{\Omega_1}\Bigg[
q_1P_y\dfrac{\partial}{\partial P_x}
\Big(\dfrac{\partial f_0/\partial P_y}{q_2P_x-z_2}\Big)+
\dfrac{\Omega_1-q_1P_x}{q_2P_x-z_2}\cdot
\dfrac{\partial^2 f_0}{\partial P_y^2}\Bigg]+
$$
$$
+\dfrac{1}{\Omega_2}\Bigg[q_2P_y\dfrac{\partial}{\partial P_x}
\Big(\dfrac{\partial f_0/\partial P_y}{q_1P_x-z_1}\Big)+
\dfrac{\Omega_2-q_2P_x}{q_1P_x-z_1}\cdot
\dfrac{\partial^2 f_0}{\partial P_y^2}
\Bigg]\Bigg\}\dfrac{1}{qP_x-z'}.
\eqno{(3.8)}
$$

Здесь
$$
z'=\dfrac{\Omega_1+\Omega_2+iy}{2}=
\dfrac{\omega_1+\omega_2+i \nu}{2k_Tv_T}=\dfrac{z_1'+z_2'}{2},
$$
$$
q=\dfrac{q_1+q_2}{2}=\dfrac{k_1+k_2}{2k_T},
$$\medskip
$$
B_0=\dfrac{1}{\pi \hat f_0(\alpha)}\int \psi_0 d^3P.
\eqno{(3.9)}
$$

Для нахождения $B_j$ подставим (3.6) в (3.7). Получим уравнение
$$
B_j\Big(\pi \hat f_0(\alpha)+
\dfrac{iy}{2}\int \dfrac{g(P)d^3P}{q_jP_x-z_j'}\Big)=
-\dfrac{e^2}{2k_T^2p_T^2v_T^2}\cdot\dfrac{1}{\Omega_j}
\int \dfrac{\Xi_{jj}({\bf P})d^3P}{q_jP_x-z_j'}.
$$

Здесь
$$
\Xi_{jj}({\bf P})=q_jP_y\dfrac{\partial}{\partial P_x}
\Big(\dfrac{\partial f_0/\partial P_y}{q_jP_x-z_j}\Big)+
\dfrac{\Omega_j-q_jP_x}{q_jP_x-z_j}\cdot
\dfrac{\partial^2 f_0}{\partial P_y^2}.
$$

Обозначим далее:

$$
J_j=\int\dfrac{g(P)d^3P}{q_jP_x-z_j'},\qquad
J_{jj}=\int\dfrac{\Xi_{jj}({\bf P})d^3P}{q_jP_x-z_j'}.
$$

Из предыдущего уравнения находим:
$$
B_j=-\dfrac{e^2}{2k_T^2p_T^2v_T^2}\cdot \gamma_j,
$$
где
$$
\gamma_j=\dfrac{\Omega_j^{-1}J_{jj}}{\pi\hat f_0(\alpha)+
\dfrac{iy}{2}J_j}.
\eqno{(3.10)}
$$

Согласно (3.6) теперь получаем:
$$
\psi_j=\dfrac{e^2}{2k_T^2p_T^2v_T^2}\Bigg[\dfrac{iy}{2}\gamma_j
\dfrac{g(P)}{q_jP_x-z_j'}-\dfrac{1}{\Omega_j}
\dfrac{\Xi_{jj}({\bf P})}{q_jP_x-z_j'}\Bigg].
\eqno{(3.11)}
$$

Для нахождения $B_0$ подставим (3.8) в (3.9). Получим уравнение
$$
B_0\Big(\pi\hat f_0(\alpha)+\dfrac{iy}{2}\int \dfrac{g(P)d^3P}
{qP_x-z'}\Big)=
$$
$$
=-\dfrac{e^2}{2k_T^2p_T^2v_T^2}
\Bigg[\dfrac{1}{\Omega_1}\int\dfrac{\Xi_{12}({\bf P})d^3P}{qP_x-z'}+
\dfrac{1}{\Omega_2}\int\dfrac{\Xi_{21}({\bf
P})d^3P}{qP_x-z'}\Bigg].
$$
Здесь
$$
\Xi_{12}({\bf P})=q_1P_y\dfrac{\partial}{\partial P_x}
\Big(\dfrac{\partial f_0/\partial P_y}{q_2P_x-z_2}\Big)+
\dfrac{\Omega_1-q_1P_x}{q_2P_x-z_2}\cdot
\dfrac{\partial^2 f_0}{\partial P_y^2},
$$
$$
\Xi_{21}({\bf P})=q_2P_y\dfrac{\partial}{\partial P_x}
\Big(\dfrac{\partial f_0/\partial P_y}{q_1P_x-z_1}\Big)+
\dfrac{\Omega_2-q_2P_x}{q_1P_x-z_1}\cdot
\dfrac{\partial^2 f_0}{\partial P_y^2}.
$$

Обозначим
$$
J_0=\int \dfrac{g(P)d^3P}{qP_x-z'},
$$
и
$$
J_{12}=\int\dfrac{\Xi_{12}({\bf P})d^3P}{qP_x-z'}, \qquad
J_{21}=\int\dfrac{\Xi_{21}({\bf P})d^3P}{qP_x-z'}.
$$

Теперь из предыдущего уравнения находим:
$$
B_0=-\dfrac{e^2}{2k_T^2p_T^2v_T^2}\cdot \gamma_0,
$$
где
$$
\gamma_0=\dfrac{\Omega_1^{-1}J_{12}+
\Omega_2^{-1}J_{21}}{\pi\hat f_0(\alpha)+\dfrac{iy}{2}J_0}.
\eqno{(3.12)}
$$

Следовательно, функция $\psi_0$ построена:
$$
\psi_0=\dfrac{e^2}{2k_T^2p_T^2v_T^2}\Bigg[\dfrac{iy}{2}\gamma_0
\dfrac{g(P)}{qP_x-z'}-\dfrac{{\Omega_1}^{-1}\Xi_{12}({\bf P})+
{\Omega_2}^{-1}\Xi_{21}({\bf P})}{qP_x-z'}\Bigg].
\eqno{(3.13)}
$$

Функция распределения во втором приближении по полю построена и
определяется равенством (2.1), в котором функция $f_1$
определяется равенством (2.9), а функция $f_2$ определяется
равенством (3.3), в котором функция $\psi_j$ определяется
равенством (3.11), а функция $\psi_0$ -- равенством (3.13).

Выпишем функцию $f_2$ в явном виде:
$$
f_2=\dfrac{e^2}{2k_T^2p_T^2v_T^2}\Bigg[\sum\limits_{j=1}^{2}E_j^2
\Bigg(\dfrac{iy}{2}\gamma_j\dfrac{g(P)}{q_jP_x-z_j'}-
\dfrac{\Omega_j^{-1}\Xi_{jj}({\bf P})}{q_jP_x-z_j'}\Bigg)+
$$
$$
+E_1E_2\Bigg(\dfrac{iy}{2}\gamma_0\dfrac{g(P)}{qP_x-z'}-
\dfrac{\Omega_1^{-1}\Xi_{12}({\bf P})+\Omega_2^{-1}
\Xi_{21}({\bf P})}{qP_x-z'}\Bigg)\Bigg].
\eqno{(3.14)}
$$

\section{Нахождение столкновительных констант}

Формулы для вычисления констант $\gamma_j$ и $\gamma_0$,
возникающих из-за наличия интеграла столкновений, упростим,
сведя трехмерные интегралы к одномерным.

Начнем с константы $\gamma_j$. Имеем:
$$
J_{jj}=\int \dfrac{\Xi_{jj}({\bf P})d^3P}{q_jP_x-z'_j}=
$$
$$
=\int \Big[q_jP_y\dfrac{\partial}{\partial P_x}
\Big(\dfrac{\partial f_0/\partial P_y}{q_jP_x-z_j}\Big)+
\dfrac{\Omega_j-q_jP_x}{q_jP_x-z_j}\cdot
\dfrac{\partial^2 f_0}{\partial
P_y^2}\Big]\dfrac{d^3P}{q_jP_x-z'_j}.
$$

Заметим, что внутренний интеграл по переменной $P_y$  от второго
слагаемого равен нулю:
$$
\int\limits_{-\infty}^{\infty}\dfrac{\partial^2f_0}{\partial P_y^2}dP_y
=\dfrac{\partial f_0}{\partial P_y}\Bigg|_{P_y=-\infty}^{P_y=+\infty}=0.
$$

Следовательно, интеграл $J_{jj}$ равен:

$$
J_{jj}=q_j\int P_y\dfrac{\partial}{\partial P_x}
\Big(\dfrac{\partial f_0/\partial P_y}{q_jP_x-z_j}\Big)
\dfrac{d^3P}{q_jP_x-z'_j}.
$$

Внутренний интеграл по переменной $P_x$ вычислим по частям:
$$
\int\limits_{-\infty}^{\infty}\dfrac{\partial}{\partial P_x}
\Big(\dfrac{\partial f_0/\partial P_y}{q_jP_x-z_j}\Big)
\dfrac{P_xdP_x}{q_jP_x-z_j'}=q_j \int\limits_{-\infty}^{\infty}
\dfrac{[\partial f_0/\partial
P_y]dP_x}{(q_jP_x-z_j')^2(q_jP_x-z_j)}.
$$

Следовательно, интеграл $J_{jj}$ равен:
$$
J_{jj}=q_j^2\int \dfrac{P_y[\partial f_0/\partial P_y]d^3P}
{(q_jP_x-z'_j)^2(q_jP_x-z_j)}.
$$

Внутренний интеграл по переменной $P_y$  проинтегрируем по
частям:
$$
\int\limits_{-\infty}^{\infty}P_y\dfrac{\partial f_0}{\partial P_y}dP_y=
P_yf_0\Bigg|_{P_y=-\infty}^{P_y=+\infty}-
\int\limits_{-\infty}^{\infty}f_0(P)dP_y=
-\int\limits_{-\infty}^{\infty}f_0(P)dP_y.
$$

Следовательно, интеграл $J_{jj}$ равен:
$$
J_{jj}=-q_j^2\int
\dfrac{f_0(P)d^3P}{(q_jP_x-z'_j)^2(q_jP_x-z_j)}.
$$

Внутренний интеграл в плоскости $(P_y,P_z)$ вычислим в полярных
координатах:
$$
\int\limits_{-\infty}^{\infty}\int\limits_{-\infty}^{\infty}
f_0(P)dP_ydP_z=\pi\ln(1+e^{\alpha-P_x^2}).
$$

Следовательно, интеграл $J_{jj}$ равен:
$$
J_{jj}=-\pi q_j^2 \int\limits_{-\infty}^{\infty}
\dfrac{\ln(1+e^{\alpha-\tau^2})d\tau}{(q_j\tau-z'_j)^2(q_j\tau-z_j)}.
$$

Таким образом,
$$
J_{jj}=-\pi q_j^2 J^1_{jj},
$$
где
$$
 J^1_{jj}= \int\limits_{-\infty}^{\infty}
\dfrac{\ln(1+e^{\alpha-\tau^2})d\tau}{(q_j\tau-z'_j)^2(q_j\tau-z_j)}.
\eqno{(4.1)}
$$

Далее, интеграл $J_j$ сведем к одномерному:
$$
J_j=\int\dfrac{g(P)d^3P}{q_jP_x-z_j'}=\int\limits_{-\infty}^{\infty}
\dfrac{dP_x}{q_jP_x-z'_j}\int\limits_{-\infty}^{\infty}
\int\limits_{-\infty}^{\infty}g(P)dP_ydP_z=$$
$$
=\pi
\int\limits_{-\infty}^{\infty}\dfrac{f_0(\tau)d\tau}{q_j\tau-z'_j}.
$$
Найдем знаменатель выражения (3.10):
$$
\pi\hat f_0(\alpha)+\dfrac{iy}{2}J_j=
\pi \int\limits_{-\infty}^{\infty}\dfrac{q_j\tau-\Omega_j}
{q_j\tau-z'_j}f_0(\tau)d\tau=\pi J^1_j,
$$
где
$$
J^1_j=\int\limits_{-\infty}^{\infty}\dfrac{q_j\tau-\Omega_j}
{q_j\tau-z'_j}f_0(\tau)d\tau.
\eqno{(4.2)}
$$

Следовательно, столкновительные константы $\gamma_j$ согласно
(3.10) равны:
$$
\gamma_j=-\dfrac{q_j^2}{\Omega_j}\cdot \dfrac{J^1_{jj}}{J^1_j},
\qquad j=1,2.
\eqno{(4.3)}
$$

Перейдем к нахождению $\gamma_0$. Имеем:
$$
J_{12}=\int \dfrac{\Xi_{12}d^3P}{q\tau-z'}=
$$
$$
=\int \Big[q_1P_y\dfrac{\partial}{\partial P_x}
\Big(\dfrac{\partial f_0/\partial P_y}{q_2P_x-z_2}\Big)+
\dfrac{\Omega_1-q_1P_x}{q_2P_x-z_2}\cdot
\dfrac{\partial^2 f_0}{\partial
P_y^2}\Big]\dfrac{d^3P}{q\tau-z'}=
$$
$$
=q_1\int P_y\dfrac{\partial}{\partial P_x}
\Big(\dfrac{\partial f_0/\partial P_y}{q_2P_x-z_2}\Big)
\dfrac{d^3P}{q\tau-z'}=
$$
$$
=q_1q\int \dfrac{P_y[\partial f_0/\partial P_y]d^3P}
{(qP_x-z')^2(q_2P_x-z_2)}=-q_1q\int\dfrac{f_0(P)d^3P}
{(qP_x-z')^2(q_2P_x-z_2)}=
$$
$$
=-\pi q_1qJ^1_{12},
$$
где
$$
J^1_{12}=\int\limits_{-\infty}^{\infty}
\dfrac{\ln(1+e^{\alpha-\tau^2})d\tau}{(q\tau-z')^2(q_2\tau-z_2)}.
\eqno{(4.4)}
$$

Аналогично,
$$
J_{21}=-\pi q_2q J^1_{21},
$$
где
$$
 J^1_{21}=\int\limits_{-\infty}^{\infty}
\dfrac{\ln(1+e^{\alpha-\tau^2})d\tau}{(q\tau-z')^2(q_1\tau-z_1)}.
\eqno{(4.5)}
$$

Знаменатель (3.12) равен:
$$
J_0=\int \dfrac{g(P)d^3P}{qP_x-z'}=\pi \int\limits_{-\infty}^{\infty}
\dfrac{f_0(\tau)d\tau}{q\tau-z'}.
$$
Теперь
$$
\pi\hat f_0(\alpha)+\dfrac{iy}{2}J_0=\pi J^1_0.
$$
Здесь
$$
J^1_0=\int\limits_{-\infty}^{\infty}
\dfrac{q\tau-\Omega}{q\tau-z'}f_0(\tau)d\tau,
\eqno{(4.6)}
$$
где
$$
\Omega=\dfrac{\Omega_1+\Omega_2}{2}.
$$
Следовательно, константа $\gamma_0$ равна:
$$
\gamma_0=-\dfrac{q}{J^1_0}\Big(\dfrac{q_1}{\Omega_1}J^1_{12}+
\dfrac{q_2}{\Omega_2}J^1_{21}\Big).
\eqno{(4.7)}
$$

\section{Плотность поперечного тока}

Найдем плотность электрического тока
$$
\mathbf{j}=e\int \mathbf{v}f \dfrac{2d^3p}{(2\pi\hbar)^3}.
\eqno{(5.1)}
$$

Во втором приближении плотность электрического тока равна:
$$
{\bf j}=e\int {\bf v}
[f_0(P)+f_1+f_2]\dfrac{2d^3p}{(2\pi\hbar)^3}.
$$

Это равенство  представим в виде суммы двух слагаемых:
$$
{\bf j}=\dfrac{2e}{(2\pi\hbar)^3}\Big[\int f_1d^3p+
\int f_2 d^3p\Big]={\bf j}^{\rm linear}+{\bf j}^{\rm quadr}.
$$

Здесь
$$
{\bf j}^{\rm linear}=\dfrac{2e}{(2\pi\hbar)^3}\int {\bf
v}f_1 d^3p=\dfrac{2e}{(2\pi\hbar)^3}
\int v_yf_1d^3p={j}_y^{\rm linear}\cdot
(0,1,0),
$$
где
$$
{j}_y^{\rm linear}=\dfrac{2e}{(2\pi\hbar)^3}\int v_yf_1d^3p,
$$
а также
$$
{\bf j}^{\rm quadr}=\dfrac{2e}{(2\pi\hbar)^3}\int {\bf v}f_2 d^3p=
{j}_x^{\rm quadr}(1,0,0),
$$
где
$$
{j}_x^{\rm quadr}=\dfrac{2e}{(2\pi\hbar)^3}\int {v}_xf_2 d^3p.
$$

Таким образом, плотность электрического тока в плазме имеет следующую
структуру:
$$
{\bf j}=(j_x^{\rm quadr}, j_y^{\rm linear},0).
$$

Из этого равенства видно, что поперечный ток, ортогональный
волновому вектору, определяется только первым, линейным
приближением, а продольный ток, направленный вдоль волнового
вектора, определяется квадратичным приближением функции
распределения.

Далее вектор плотности тока будем обозначать через
$$
\mathbf{j}=(j_x,j_y,0).
$$

Здесь $j_y$ -- плотность поперечного тока. Согласно (5.1) имеем:
$$
j_y=e\int v_yf_1 \dfrac{2d^3p}{(2\pi\hbar)^3}=
\dfrac{2ep_T^3v_T}{(2\pi\hbar)^3}\int f_1P_yd^3P.
\eqno{(5.2)}
$$

Этот ток направлен вдоль напряженности электромагнитных полей,
его плотность определяется только первым приближением
функции распределения (2.9).

Второе приближение функции распределения вклад в плотность тока
не вносит.

Подставим (2.9) в (5.2). Получаем, что
$$
j_x=\dfrac{2ie^2p_T^2}{(2\pi\hbar)^3k_T}\sum\limits_{j=1}^{2}E_j
\int \dfrac{P_y[\partial f_0/\partial P_y]}{q_jP_x-z_j}d^3P.
\eqno{(5.3)}
$$

Этот ток состоит из двух слагаемых, пропорциональных
первой степени  величин напряженностей электрических полей.

Упростим эту формулу. Заметим, что внутренний интеграл по
переменной $P_y$ равен:
$$
\int\limits_{-\infty}^{\infty}P_y\dfrac{\partial f_0(P)}{\partial P_y}dP_y=
P_yf_0(P)\Bigg|_{P_y=-\infty}^{P_y=+\infty}-\int\limits_{-\infty}^{\infty}
f_0(P)dP_y=-\int\limits_{-\infty}^{\infty}
f_0(P)dP_y.
$$

Следовательно, плотность поперечного тока согласно (5.3) равна:
$$
j_y=-\dfrac{2ie^2p_T^2}{(2\pi\hbar)^3k_T}
\sum\limits_{j=1}^{2}E_j\int\dfrac{f_0(P)d^3P}{q_jP_x-z_j}.
$$

Вычислим внутренний двойной интеграл в плоскости $(P_y,P_z)$.
Тогда получаем, что
$$
j_y=-\dfrac{2i\pi e^2p_T^2}{(2\pi\hbar)^3k_T}\sum\limits_{j=1}^{2}E_j
\int\limits_{-\infty}^{\infty}\dfrac{\ln(1+e^{\alpha-\tau^2})d\tau}
{q_j\tau-z_j}.
\eqno{(5.4)}
$$

Найдем числовую плотность концентрацию частиц плазмы, отвечающую
абсолютному распределению Ферми---Дирака
$$
N=\int f_0(P)\dfrac{2d^3p}{(2\pi\hbar)^3}=
\dfrac{8\pi p_T^3}{(2\pi\hbar)^3}\int\limits_{0}^{\infty}
\dfrac{e^{\alpha-P^2}P^2dP}{1+e^{\alpha-P^2}}=
\dfrac{k_T^3}{2\pi^2}l_0(\alpha),
$$

Воспользуемся связью числовой плотности в равновесном состоянии,
теплового волнового числа и безразмерного химического
потенциала:
$$
N=\dfrac{k_T^3}{2\pi^2}l_0(\alpha),
\eqno{(5.5)}
$$
где
$$
l_0(\alpha)=\int\limits_{0}^{\infty}\ln(1+e^{\alpha-\tau^2})d\tau.
$$

Теперь плотность тока (5.4) можно представить в виде:
$$
j_x=-iy\sigma_0\dfrac{1}{2l_0(\alpha)}\sum\limits_{j=1}^{2}E_j
\int\limits_{-\infty}^{\infty}\dfrac{\ln(1+e^{\alpha-\tau^2})}
{q_j\tau-z_j}d\tau,
$$
или, короче,
$$
j_x=\sigma_0(j_1E_1+j_2E_2),
\eqno{(5.6)}
$$
где $\sigma_0=e^2N/m\nu$ -- статическая электрическая
проводимость,
$$
j_j=-\dfrac{iy}{2l_0(\alpha)}\int\limits_{-\infty}^{\infty}
\dfrac{\ln(1+e^{\alpha-\tau^2})}{q_j\tau-z_j}d\tau.
$$

Формулу (5.6) представим в бескоординатном (инвариантном) виде:
$$
{\bf j}=\sigma_0(j_1{\bf E}_1+j_2{\bf E}_2).
\eqno{(5.7)}
$$

В формуле (5.7)
$$
j_j=-\dfrac{iy}{2l_0(\alpha)}\int\limits_{-\infty}^{\infty}
\dfrac{\ln(1+e^{\alpha-\tau^2})}{{\bf q}_j{\bf P}-z_j}d\tau,
$$
где
$$
\tau=\dfrac{{\bf q}_j{\bf P}}{q_j}=\dfrac{{\bf k}_j{\bf
P}}{k_j}, \qquad j=1,2.
$$

Выделим физически значимую действительную часть поперечного
тока. Используя формулу (5.6) находим, что
$$
\Re j_x=\sigma_0\dfrac{y}{2l_0(\alpha)}\int\limits_{-\infty}^{\infty}
\ln(1+e^{\alpha-\tau^2})\times $$$$ \times\sum\limits_{j=1}^{2}E_{0j}
\dfrac{y\cos (k_jx-\omega_j t)+(q_j\tau-\omega_j t)
\sin(k_jx-\omega_j t)}{|q_j\tau-\Omega_j-iy|^2}d\tau.
\eqno{(5.8)}
$$

Здесь $E_{0j}$ -- действительные амплитуды напряженностей
электрических полей.

Формулу (5.8) также представим в бескоординатном виде:
$$
\Re {\bf j}=\sigma_0\dfrac{y}{2l_0(\alpha)}\int\limits_{-\infty}^{\infty}
\ln(1+e^{\alpha-\tau^2})\times $$$$ \times\sum\limits_{j=1}^{2}E_{0j}
\dfrac{y\cos ({\bf k}_j{\bf r}-\omega_j t)+({\bf q}_j{\bf P}-\omega_j t)
\sin({\bf k}_j{\bf r}-\omega_j t)}{|{\bf q}_j{\bf P}-\Omega_j-iy|^2}d\tau.
$$

\section{ПЛОТНОСТЬ ПРОДОЛЬНОГО ТОКА}

Далее будем исследовать продольный ток $j_x^{\rm quadr}$,
имеющий второй порядок малости по величинам напряженностей
электрических полей. Будем обозначать этот ток $j_x$. Имеем:
$$
j_x=e\int v_xf_2\dfrac{2d^3p}{(2\pi\hbar)^3}=
\dfrac{2ep_T^3v_T}{(2\pi\hbar)^3}\int P_xf_2d^3P.
$$

С помощью разложения (3.12) представим продольный ток в виде трех
слагаемых:
$$
j_x=E_1^2j_1+E_2^2j_2+E_1E_2j_0.
\eqno{(6.1)}
$$

Здесь
$$
j_j=\dfrac{2e^3p_Tv_T}{(2\pi \hbar)^3}\int P_x\psi_jd^3P,\qquad
j=0,1,2.
$$

Представим предыдущие три равенства в явном виде, воспользовавшись
связью числовой плотности с тепловым волновым числом и
безразмерным химическим потенциалом (5.5). В результате
получаем:
$$
j_j=\dfrac{e^3N}{4\pi l_0(\alpha)k_T^2p_T^2v_T}\cdot \Bigg[
\dfrac{iy}{2}\gamma_j\int \dfrac{g(P)P_xd^3P}{q_jP_x-z'_j}-
\dfrac{1}{\Omega_j}\int\dfrac{\Xi_{jj}({\bf
P})P_x}{q_jP_x-z'_j}d^3P\Bigg],\;j=1,2,
\eqno{(6.2)}
$$

$$
j_0=\dfrac{e^3N}{4\pi l_0(\alpha)k_T^2p_T^2v_T}\cdot \Bigg[
\dfrac{iy}{2}\gamma_0\int \dfrac{g(P)P_xd^3P}{qP_x-z'}-
\dfrac{1}{\Omega_1}\int\dfrac{\Xi_{12}({\bf
P})P_x}{qP_x-z'}d^3P-
$$
$$
-\dfrac{1}{\Omega_2}\int\dfrac{\Xi_{21}({\bf P})P_x}{qP_x-z'}d^3P\Bigg].
\eqno{(6.3)}
$$

Вычислим  интегралы:
$$
\int \dfrac{g(P)P_xd^3P}{q_jP_x-z'_j}=\pi
\int\limits_{-\infty}^{\infty}
\dfrac{f_0(P)P_xdP_x}{q_jP_x-z'_j}=-\dfrac{\pi q_j}{2}
\int\limits_{-\infty}^{\infty}
\dfrac{\ln(1+e^{\alpha-\tau^2})d\tau}{(q_j\tau-z'_j)^2},
$$
и
$$
\int \dfrac{g(P)P_xd^3P}{qP_x-z'}=\pi
\int\limits_{-\infty}^{\infty}
\dfrac{f_0(P)P_xdP_x}{qP_x-z'}=-\dfrac{\pi q}{2}
\int\limits_{-\infty}^{\infty}
\dfrac{\ln(1+e^{\alpha-\tau^2})d\tau}{(q\tau-z')^2}.
$$

Далее
$$
\int \dfrac{\Xi_{jj}({\bf P})P_xd^3P}{q_jP_x-z'_j}=$$$$=
\int \Bigg[q_jP_y\dfrac{\partial}{\partial P_x}
\Big(\dfrac{\partial f_0/\partial P_y}{q_jP_x-z_j}\Big)+
\dfrac{\Omega_j-q_jP_x}{q_jP_x-z_j}\dfrac{\partial^2 f_0}{\partial P_y^2}
\Bigg]\dfrac{P_xd^3P}{q_jP_x-z_j'}=
$$
$$
=q_j\int P_y\dfrac{\partial}{\partial P_x}
\Big(\dfrac{\partial f_0/\partial P_y}{q_jP_x-z_j}\Big)
\dfrac{P_xd^3P}{q_jP_x-z_j'}=
$$
$$
=q_jz'_j\int\dfrac{P_y[\partial f_0/\partial P_y]d^3P}
{(q_jP_x-z_j)(q_jP_x-z'_j)^2}=-q_jz'_j\int \dfrac{f_0(P)d^3P}
{(q_jP_x-z_j)(q_jP_x-z'_j)^2}=
$$
$$
=-\pi q_jz'_j\int\limits_{-\infty}^{\infty}
\dfrac{\ln(1+e^{\alpha-\tau^2})d\tau}{(q_j\tau-z_j)(q_j\tau-z'_j)^2}.
$$

Аналогично,
$$
\int\dfrac{\Xi_{12}({\bf P})P_x}{qP_x-z'}d^3P=
-\pi q_1z'\int\limits_{-\infty}^{\infty}
\dfrac{\ln(1+e^{\alpha-\tau^2})d\tau}{(q_2\tau-z_2)(q\tau-z')^2}
$$
и
$$
\int\dfrac{\Xi_{21}({\bf P})P_x}{qP_x-z'}d^3P=
-\pi q_2z'\int\limits_{-\infty}^{\infty}
\dfrac{\ln(1+e^{\alpha-\tau^2})d\tau}{(q_1\tau-z_1)(q\tau-z')^2}.
$$

В предыдущих равенствах были использованы соотношения:
$$
\int\limits_{-\infty}^{\infty}\dfrac{\partial^2f_0}{\partial P_y^2}dP_y
=\dfrac{\partial f_0}{\partial P_y}\Bigg|_{P_y=-\infty}^{P_y=+\infty}=0.
$$
и
$$
q_1\int\limits_{-\infty}^{\infty}P_y\dfrac{\partial}{\partial P_x}
\Big(\dfrac{\partial f_0/\partial P_y}{q_jP_x-z_j}\Big)
\dfrac{P_xdP_x}{q_jP_x-z'_j}=q_1z_j' \int\limits_{-\infty}^{\infty}
\dfrac{P_y[\partial f_0/\partial P_y]dP_x}{(q_jP_x-z_j')^2(q_jP_x-z_j)},
$$
где $j=1,2$,
$$
q_1\int\limits_{-\infty}^{\infty}P_y\dfrac{\partial}{\partial P_x}
\Big(\dfrac{\partial f_0/\partial P_y}{q_2P_x-z_2}\Big)
\dfrac{P_xdP_x}{qP_x-z'}=-q_1z'\int\limits_{-\infty}^{\infty}
\dfrac{P_y[\partial f_0/\partial P_y]dP_x}{(q_2P_x-z_2)(qP_x-z')},
$$
$$
q_2\int\limits_{-\infty}^{\infty}P_y\dfrac{\partial}{\partial P_x}
\Big(\dfrac{\partial f_0/\partial P_y}{q_1P_x-z_1}\Big)
\dfrac{P_xdP_x}{qP_x-z'}=-q_2z'\int\limits_{-\infty}^{\infty}
\dfrac{P_y[\partial f_0/\partial P_y]dP_x}{(q_1P_x-z_1)(qP_x-z')},
$$
$$
\int\limits_{-\infty}^{\infty}P_y\dfrac{\partial f_0}{\partial P_y}dP_y=
P_yf_0\Bigg|_{P_y=-\infty}^{P_y=+\infty}-
\int\limits_{-\infty}^{\infty}f_0(P)dP_y=
-\int\limits_{-\infty}^{\infty}f_0(P)dP_y,
$$
и
$$
\int\limits_{-\infty}^{\infty}\int\limits_{-\infty}^{\infty}
f_0(P)dP_ydP_z=\pi\ln(1+e^{\alpha-P_x^2}).
$$

Теперь  равенства (6.2) и (6.3) сводятся к одномерным интегралам:
$$
j_j=\dfrac{e^3N}{4l_0(\alpha)k_T^2p_T^2v_T}\Bigg[
-\dfrac{iyq_j}{4}\gamma_j\int\limits_{-\infty}^{\infty}
\dfrac{\ln(1+e^{\alpha-\tau^2})d\tau}{(q_j\tau-z'_j)^2}+$$$$+
\dfrac{q_jz'_j}{\Omega_j}\int\limits_{-\infty}^{\infty}
\dfrac{\ln(1+e^{\alpha-\tau^2})d\tau}{(q_j\tau-z_j)(q_j\tau-z'_j)^2}
\Bigg]
\eqno{(6.4)}
$$
и
$$
j_0=\dfrac{e^3N}{4l_0(\alpha)k_T^2p_T^2v_T}\Bigg[
-\dfrac{iyq}{4}\gamma_0\int\limits_{-\infty}^{\infty}
\dfrac{\ln(1+e^{\alpha-\tau^2})d\tau}{(q\tau-z')^2}+$$$$+
\dfrac{q_1z'}{\Omega_1}\int\limits_{-\infty}^{\infty}
\dfrac{\ln(1+e^{\alpha-\tau^2})d\tau}{(q_2\tau-z_2)(q\tau-z')^2}+
\dfrac{q_2z'}{\Omega_2}\int\limits_{-\infty}^{\infty}
\dfrac{\ln(1+e^{\alpha-\tau^2})d\tau}{(q_1\tau-z_1)(q\tau-z')^2}
\Bigg].
\eqno{(6.5)}
$$

Перепишем равенства (6.4) и (6.5) с помощью ранее введенных
обозначений (4.1), (4.2), (4.4), (4.5) и (4.6) из п. 4:
$$
j_j=\dfrac{e^3N}{4l_0(\alpha)k_T^2p_T^2v_T}\Bigg[
-\dfrac{iyq_j}{4}\gamma_j J_j^2+
\dfrac{q_jz'_j}{\Omega_j}J^1_{jj}\Bigg]
\eqno{(6.6)}
$$
и
$$
j_0=\dfrac{e^3N}{4l_0(\alpha)k_T^2p_T^2v_T}\Bigg[
-\dfrac{iyq}{4}\gamma_0J_0^2+
\dfrac{q_1z'}{\Omega_1}J^1_{12}+
\dfrac{q_2z'}{\Omega_2}J^1_{21}\Bigg].
\eqno{(6.7)}
$$

В равенствах (6.6) и (6.7) введены обозначения:
$$
J_j^2=\int\limits_{-\infty}^{\infty}
\dfrac{\ln(1+e^{\alpha-\tau^2})d\tau}{(q_j\tau-z'_j)^2}
$$
и
$$
J_0^2=\int\limits_{-\infty}^{\infty}
\dfrac{\ln(1+e^{\alpha-\tau^2})d\tau}{(q\tau-z')^2}.
$$

Теперь воспользуемся выражениями столкновительных констант (4.3)
и  (4.7). В результате получаем следующие выражения для
составляющих продольного электрического тока:
$$
j_j=\dfrac{e^3N}{4l_0(\alpha)k_T^2p_T^2v_T}
\dfrac{q_j J_{jj}^1}{\Omega_j}\Bigg[z_j'+
\dfrac{iy}{4}q_j^2\dfrac{J_j^2}{J_j^1}\Bigg]
\eqno{(6.8)}
$$
и
$$
j_0=\dfrac{e^3N}{4l_0(\alpha)k_T^2p_T^2v_T}\Bigg[
z'+\dfrac{iy}{4}q^2\dfrac{J_0^2}{J_0^1}\Bigg]
\Bigg[\dfrac{q_1}{\Omega_1}J^1_{12}+\dfrac{q_2}{\Omega_2}J^1_{21}\Bigg].
\eqno{(6.9)}
$$

В выражении перед интегралами из (6.8) и (6.9) выделим плазменную
(ленгмюровскую) частоту
$$
\omega_p=\sqrt{\dfrac{4\pi e^2N}{m}}.
$$

Получим
$$
\dfrac{e^3N}{4l_0(\alpha)k_T^2p_T^2v_T}=
\dfrac{\sigma_{l,tr}k_T}{16\pi l_0(\alpha)}.
$$

Здесь $\sigma_{l,tr}$ -- продольно-поперечная проводимость,
$$
\sigma_{l,tr}=\dfrac{e\hbar}{p_T^2}
\Big(\dfrac{\hbar \omega_p}{mv_T^2}\Big)^2=
\dfrac{e}{k_Tp_T}\Big(\dfrac{\omega_p}{k_Tv_T}\Big)^2=
\dfrac{e\Omega_p^2}{p_Tk_T},
$$
$$
\Omega_p=\dfrac{\omega_p}{k_Tv_T}=\dfrac{\hbar\omega_p}{mv_T^2}
$$
-- безразмерная плазменная частота.

Теперь составляющие тока равны:
$$
j_j=\dfrac{\sigma_{l,tr}k_jJ_{jj}}{16\pi l_0(\alpha)\Omega_j}
\Big[\Omega_j+\dfrac{iy}{2}\Big(1+q_j^2\dfrac{J_j^2}{2J_j^1}\Big)\Big],
\eqno{(6.10)}
$$
$$
j_0=\dfrac{\sigma_{l,tr}}{16\pi l_0(\alpha)}
\Big(k_1\dfrac{J^1_{12}}{\Omega_1}+k_2\dfrac{J^1_{21}}{\Omega_2}\Big)
\times \qquad \qquad$$$$\qquad\qquad \times\Bigg\{\dfrac{\Omega_1+
\Omega_2}{2}+\dfrac{iy}{2}\Bigg[1+\Big(\dfrac{q_1+q_2}{2}\Big)^2
\dfrac{J_0^2}{2J_0^1}\Bigg]\Bigg\}.
\eqno{(6.11)}
$$

{\sc Замечание.} Из формул (6.10) и (6.11) видно, что при $y=0$ (или
$\nu=0$), т.е. когда столкновительная плазма переходит в
бесстолкновительную ($z\to x, z'\to x$),
эти формулы переходят в формулы  для бесстолкновительной плазмы:
$$
j_j=\dfrac{\sigma_{l,tr}k_j}{16\pi l_0(\alpha)}
\int\limits_{-\infty}^{\infty}
\dfrac{\ln(1+e^{\alpha-P_x^2})dP_x}{(q_jP_x-\Omega_j)^3},
$$

$$
j_0=\dfrac{\sigma_{l,tr}(\Omega_1+\Omega_2)}{16\pi
l_0(\alpha)}\Bigg[
\dfrac{k_1}{\Omega_1}\int\limits_{-\infty}^{\infty}
\dfrac{\ln(1+e^{\alpha-\tau^2})d\tau}
{[(q_1+q_2)\tau-(\Omega_1+\Omega_2)]^2(q_2\tau-\Omega_2)}+
$$
$$
+\dfrac{k_2}{\Omega_2}\int\limits_{-\infty}^{\infty}
\dfrac{\ln(1+e^{\alpha-\tau^2})d\tau}
{[(q_1+q_2)\tau-(\Omega_1+\Omega_2)]^2(q_1\tau-\Omega_1)}\Bigg].
$$

Если ввести поперечные электрические поля
$$
\mathbf{E_j}^{\bf \rm tr}=\mathbf{E_j}-\dfrac{\mathbf{k_j(E_jk_j)}}{k_j^2}=
\mathbf{E_j}-\dfrac{\mathbf{q_j(E_jq_j)}}{q_j^2},
$$
то выражение для продольного тока можно записать в инвариантой форме
$$
\mathbf{j}=\sigma_{l,tr}[J_1{\bf k}_1({\bf E}_1^{tr})^2+
J_2{\bf k}_2({\bf E}_2^{tr})+{\bf E}_1^{tr}{\bf E}_2^{tr}
(J_{12}{\bf k}_1+J_{21}{\bf k}_2)].
$$

Здесь
$$
J_j=\dfrac{1}{16\pi l_0(\alpha)}\dfrac{J^1_{jj}}{\Omega_j}\Big(z'_j+
\dfrac{iy}{4}q_j^2\dfrac{J^2_j}{J^1_j}\Big),
$$
$$
J_{12}=\dfrac{1}{16\pi l_0(\alpha)}\dfrac{J^1_{12}}{\Omega_1}
\Big(z'+\dfrac{iy}{4}q^2\dfrac{J^2_0}{J^1_0}\Big),
$$
$$
J_{21}=\dfrac{1}{16\pi l_0(\alpha)}\dfrac{J^1_{21}}{\Omega_2}
\Big(z'+\dfrac{iy}{4}q^2\dfrac{J^2_0}{J^1_0}\Big).
$$

Перейдем к рассмотрению случая малых значений волновых чисел.
В линейном приближении по $k_1$ и $k_2$ мы получаем:
$$
j_j=-\dfrac{1}{8\pi \Omega_j(\Omega_j+iy/2)(\Omega_j+iy)}
$$
$$
j_{12}=-\dfrac{2}{8\pi \Omega_1(\Omega_2+iy)(\Omega_1+\Omega_2+iy)},
$$
$$
j_{21}=-\dfrac{2}{8\pi
\Omega_2(\Omega_1+iy/2)(\Omega_1+\Omega_2+iy)}.
$$

Таким образом, плотность продольного тока в случае малых
значений волновых чисел равна:
$$
\mathbf{j}=-\dfrac{\sigma_{l,tr}}{8\pi}\Bigg[
\dfrac{{\bf k}_1({\bf E}_1^{tr})^2}{\Omega_1(\Omega_1+iy/2)(\Omega_1+iy)}+
\dfrac{{\bf k}_2({\bf E}_2^{tr})^2}{\Omega_2(\Omega_2+iy/2)(\Omega_2+iy)}+
$$
$$
+\dfrac{2{\bf E}_1^{tr}{\bf E}_2^{tr}}{\Omega_1+\Omega_2+iy}
\Big(\dfrac{{\bf k}_1}{\Omega_1(\Omega_2+iy)}+
\dfrac{{\bf k}_2}{\Omega_2(\Omega_2+iy)}\Big)\Bigg].
$$

{\sc Замечание.} При $y=0\; (\nu=0)$ из этих формул
получаем соответствующие формулы  для
продольного тока в случае
малых значений волнового числа в бесстолкновительной плазме.

\section{ЗАКЛЮЧЕНИЕ}

Проведем графическое исследование действительных и мнимых
частей  плотностей безразмерных "перекрестных"\, токов  $J_{12}$ и $J_{21}$.
В случае одного электромагнитного поля в работе \cite{Lat7} было
проведено графическое исследование безразмерной части
продольного тока. При этом исследовались составляющие
безразмерных плотностей тока вида $J_1$ (или $J_2$).
При этом использовались формулы для $J_1$ или $J_2$. Поэтому в
настоящей работе будем исследовать "перекрестные"\,
величины $J_{12}$ и $J_{21}$.

На рис. 1--4 представлено поведение действительной и мнимых
частей плотностей "перекрестных"\,токов
при значениях величины безразмерного химического потенциала
$\alpha=2, 0,-5$ в зависимости от безразмерного волнового числа $q_1$,
при этом $\Omega_1=1, \Omega_2=0.5, q_2=0.1$. При этом на рис. 1
и 2 приводится поведение действительной и мнимой частей тока $J_{12}$,
а на рис. 3 и 4 приводится поведение действительной и мнимой
частей тока $J_{21}$.

Из рис. 1 и 2 видно, что действительная часть $J_{12}$ с ростом
$q_1$ имеет сначала минимум, а затем и максимум. С ростом
степени вырождения электронного газа увеличивается область
значений и действительной части, и мнимой. При этом точки
экстремумов сдвигаются в область малых значений $q_1$.
Точно такое же поведение обнаруживает и перекрестная часть тока
$J_{21}$. Однако, с ростом величины химического потенциалау
мнимой части появляется и минимум.

На рис. 5--8 представим поведение действительных и мнимых частей
плотностей $J_{12}$ (рис. 5б6) и $J_{21}$ (рис. 7б8) при
$\Omega_1=1$, $q_2=0.1$ и при  $\Omega_2=0.1, 0.2, 0.3$
в зависимости от безразмерного волнового числа $q_1$.

Из этих рисунков видно, что действительная часть плотности
"перекрестного"\, тока имеют один минимум и один максимум,
причем с ростом $q_1$ эти величины стремятся к нулю.
С ростом частоты колебаний второго электромагнитного поля
величина максимума уменьшается, а величина минимума
увеличивается. Мнимая часть тока $J_{12}$ имеет один максимум.

С ростом волнового числа $q_1$ мнимые
части $J_{12}$ независимо от частоты колебаний второго
электромагнитного поля $\Omega_2$ сближаются и при $q_1\to \infty$
совпадают. Для плотности $J_{21}$ мнимые части всех кривых
совпадают и при малых и при больших значениях первого волнового
числа.
Отметим также, что мнимые части "перекрестных"\, токов имеют
максимум при всех значениях частоты колебаний второго
электромагнитного поля. При уменьшении частоты колебаний первого
электромагнитного поля у мнимых частей появляется минимум. При
уменьшении частоты колебаний $\Omega_1$ величина максимума
увеличивается, а минимума -- уменьшается.

В настоящей работе решена следующая задача: в плазме с произвольной
степенью вырождения электронного газа, распространяются две
электромагнитные волны с коллинеарными волновыми векторами.
Уравнение Власова решается методом последовательных приближений,
считая малыми параметрами одного порядка величины напряженностей
соответствующих электрических полей. Используется квадратичное разложение
функции распределения.

Оказалось, что учет нелинейности электромагнитных полей обнаруживает
генерирование электрического тока, ортогонального к направлению
электрического поля (т.е. направлению известного классического поперечного
электрического тока). Найдена величина поперечного и продольного
электрических токов.  Рассмотрен случай малых значений волновых чисел.
Проведено графическое исследование так называемых
"перекрестных"\, слагаемых $J_{12}$ и $J_{21}$, составляющих величину плотности
продольного электрического тока.

В дальнейшем авторы намерены рассмотреть задачи о колебаниях
плазмы и о скин-эффекте с использованием квадратичного по
потенциалу разложения функции распределения.

\clearpage

\begin{figure}[ht]\center
\includegraphics[width=16.0cm, height=10cm]{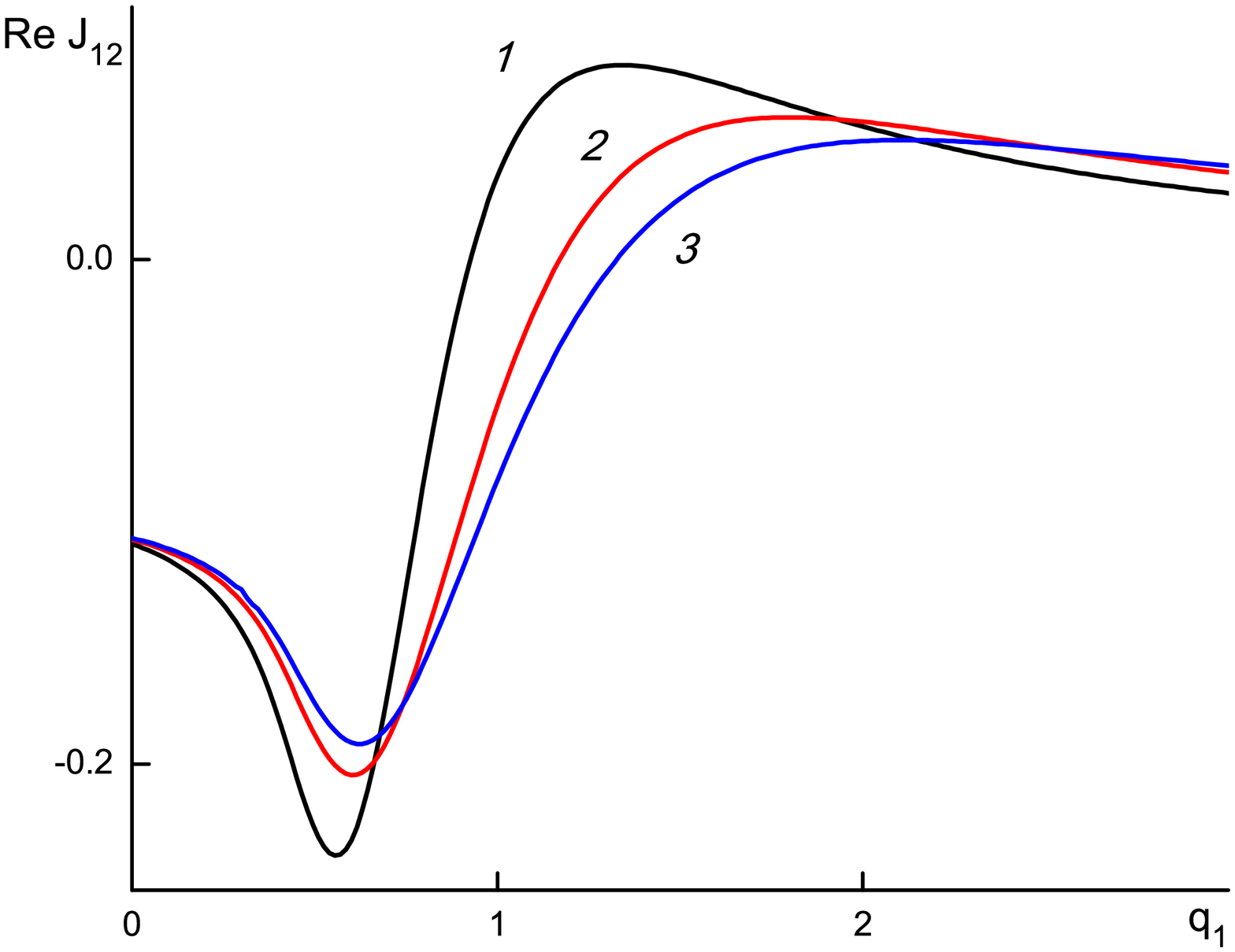}
\center{Fig. 1. Real part of dimensionless density of longitudinal
"crossed"\, current $J_{12}$, $\Omega_1=1, \Omega_2=0.5, q_2=0.1$.
Curves $1,2,3$ correspond to values of
dimensionless chemical potential $\alpha=2, 0, -5$.}
\includegraphics[width=17.0cm, height=10cm]{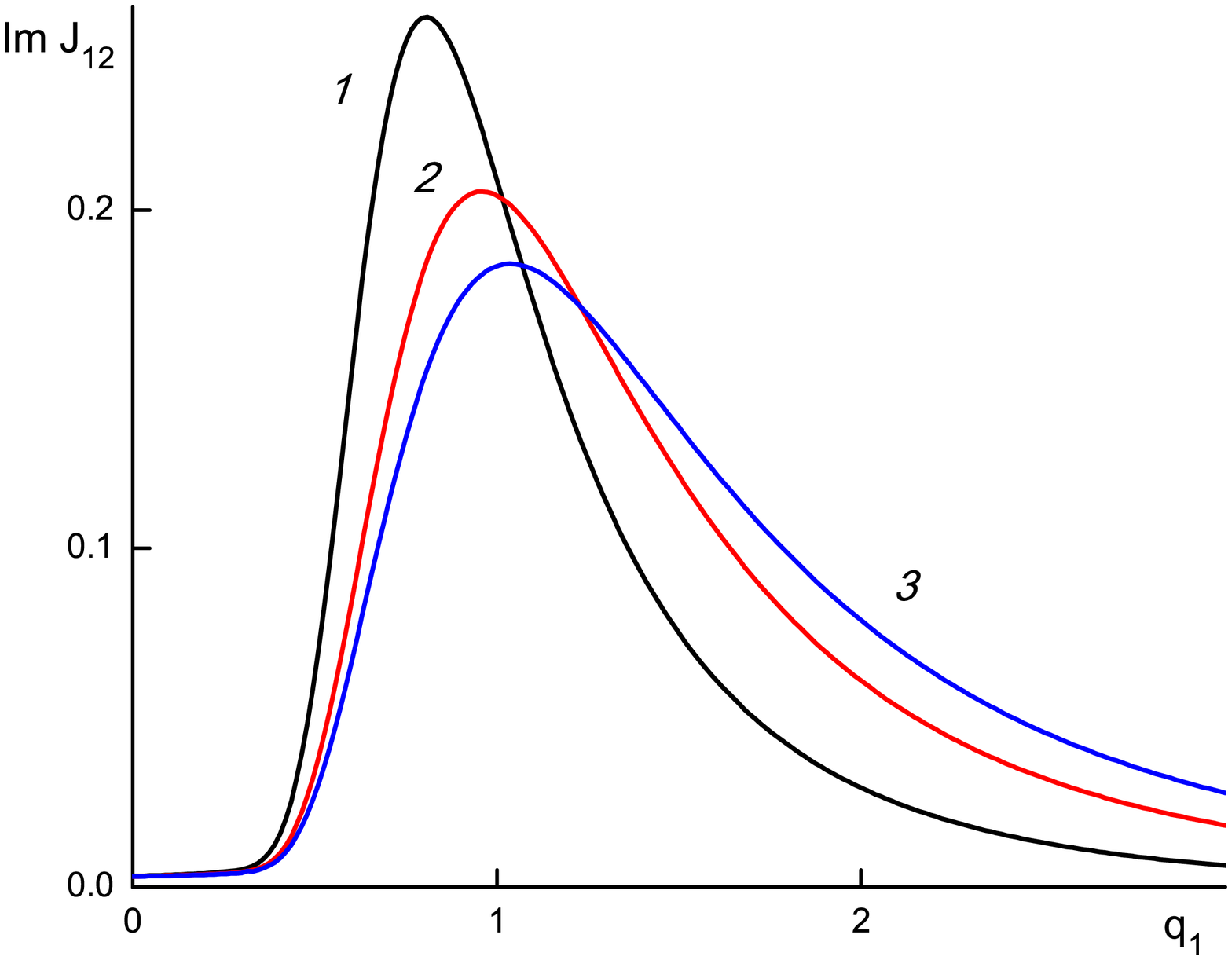}
\center{Fig. 2. Imaginary part of dimensionless density of longitudinal
"crossed"\, current $J_{12}$,  $\Omega_1=1, \Omega_2=0.5, q_2=0.1$.
Curves $1,2,3$ correspond to values of
dimensionless chemical potential $\alpha=2, 0, -5$.}
\end{figure}

\begin{figure}[ht]\center
\includegraphics[width=16.0cm, height=10cm]{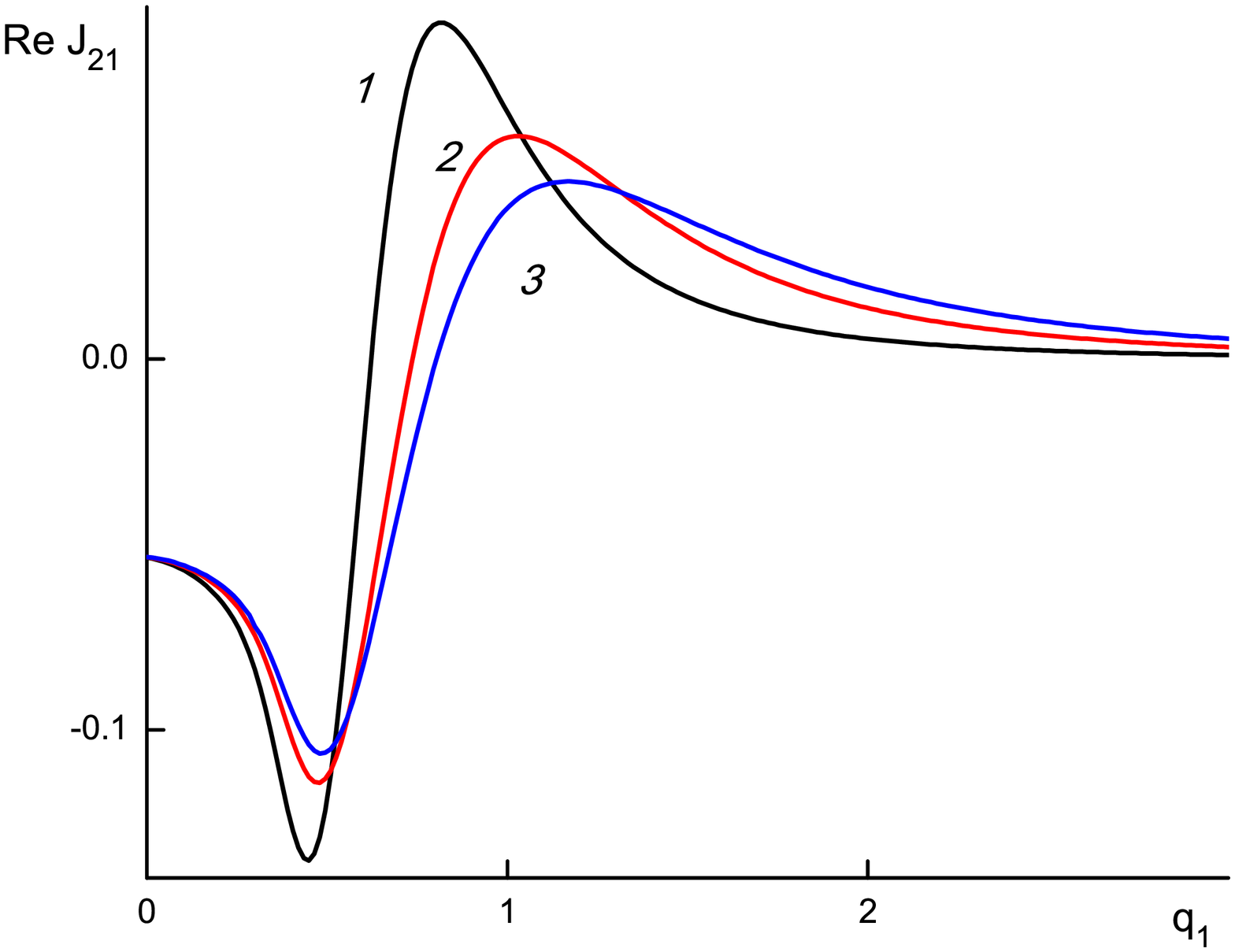}
\center{Fig. 3. Real part of dimensionless density of longitudinal
"crossed"\, current $J_{21}$, $\Omega_1=1, q_2=0.1$.
Curves $1,2,3$ correspond to values of
dimensionless oscillation frequency of second electromagnetic field
$\Omega_2=0.1, 0.2, 0.3$.}
\includegraphics[width=17.0cm, height=10cm]{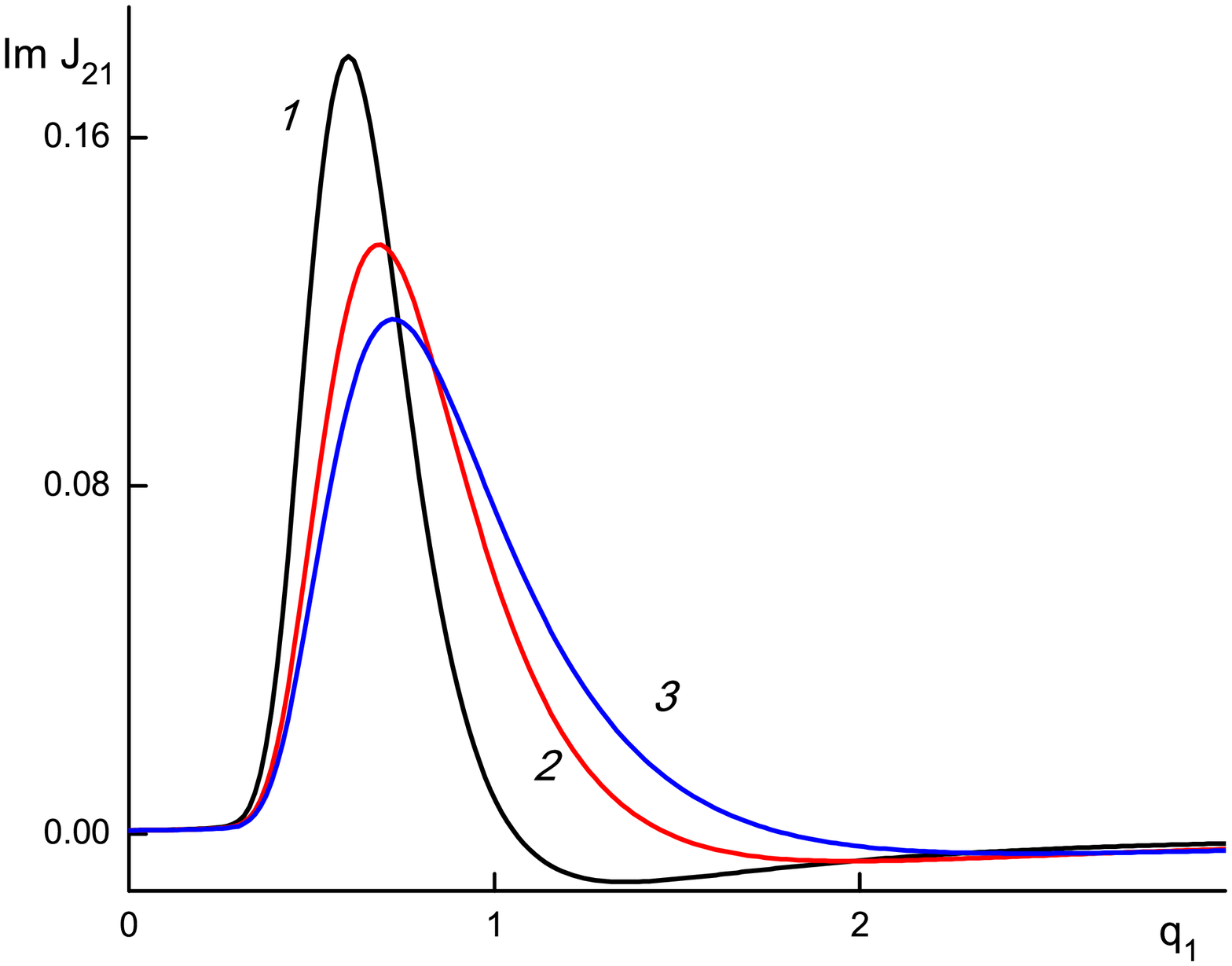}
\center{Fig. 4. Imaginary part of dimensionless density of longitudinal
"crossed"\, current $J_{21}$, $\Omega_1=1, q_2=0.1$.
Curves $1,2,3$ correspond to values of
dimensionless oscillation frequency of second electromagnetic field
$\Omega_2=0.1, 0.2, 0.3$.}
\end{figure}

\begin{figure}[ht]\center
\includegraphics[width=16.0cm, height=10cm]{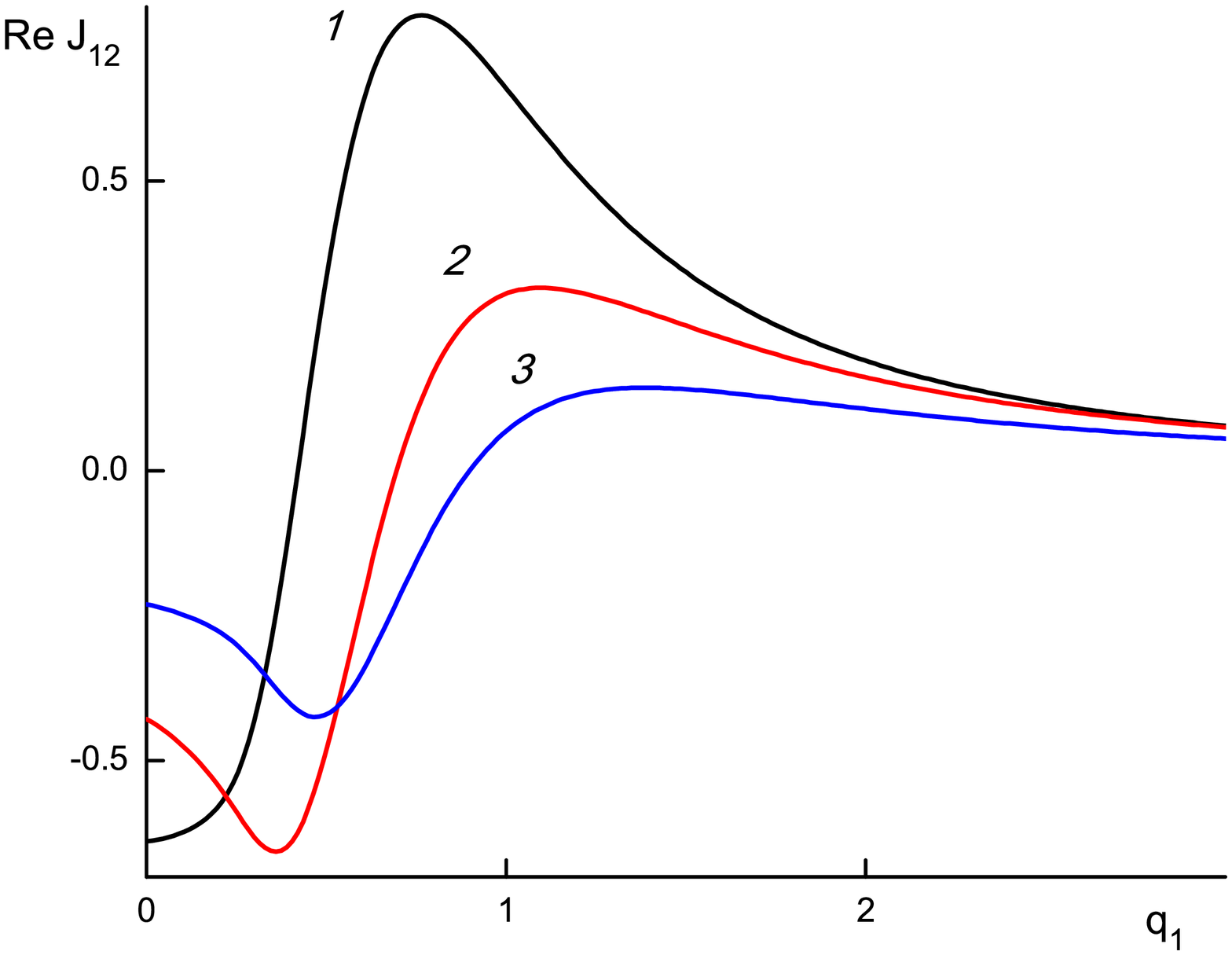}
\center{Fig. 5. Real  part of dimensionless density of longitudinal
"crossed"\, current $J_{12}$, $\Omega_1=1, \Omega_2=0.5, q_2=0.1$.
Curves $1,2,3$ correspond to values of
dimensionless oscillation frequency of second electromagnetic field
$\Omega_2=0.1, 0.2, 0.3$.}
\includegraphics[width=17.0cm, height=10cm]{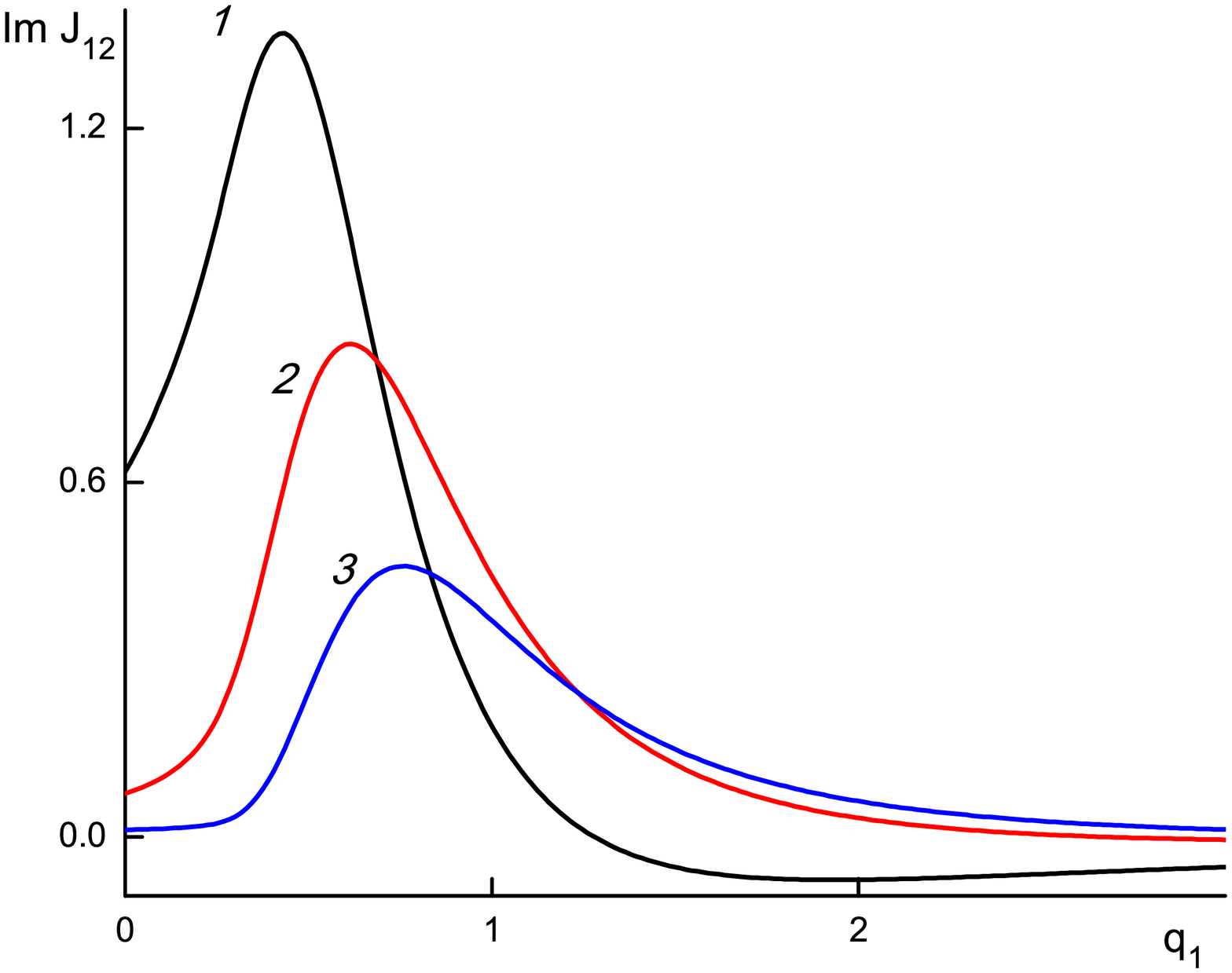}
\center{Fig. 6. Imaginary  part of dimensionless density of longitudinal
"crossed"\, current $J_{12}$, $\Omega_1=1, \Omega_2=0.5, q_2=0.1$.
Curves $1,2,3$ correspond to values of
dimensionless oscillation frequency of second electromagnetic field
$\Omega_2=0.1, 0.2, 0.3$.}
\end{figure}

\begin{figure}[ht]\center
\includegraphics[width=16.0cm, height=10cm]{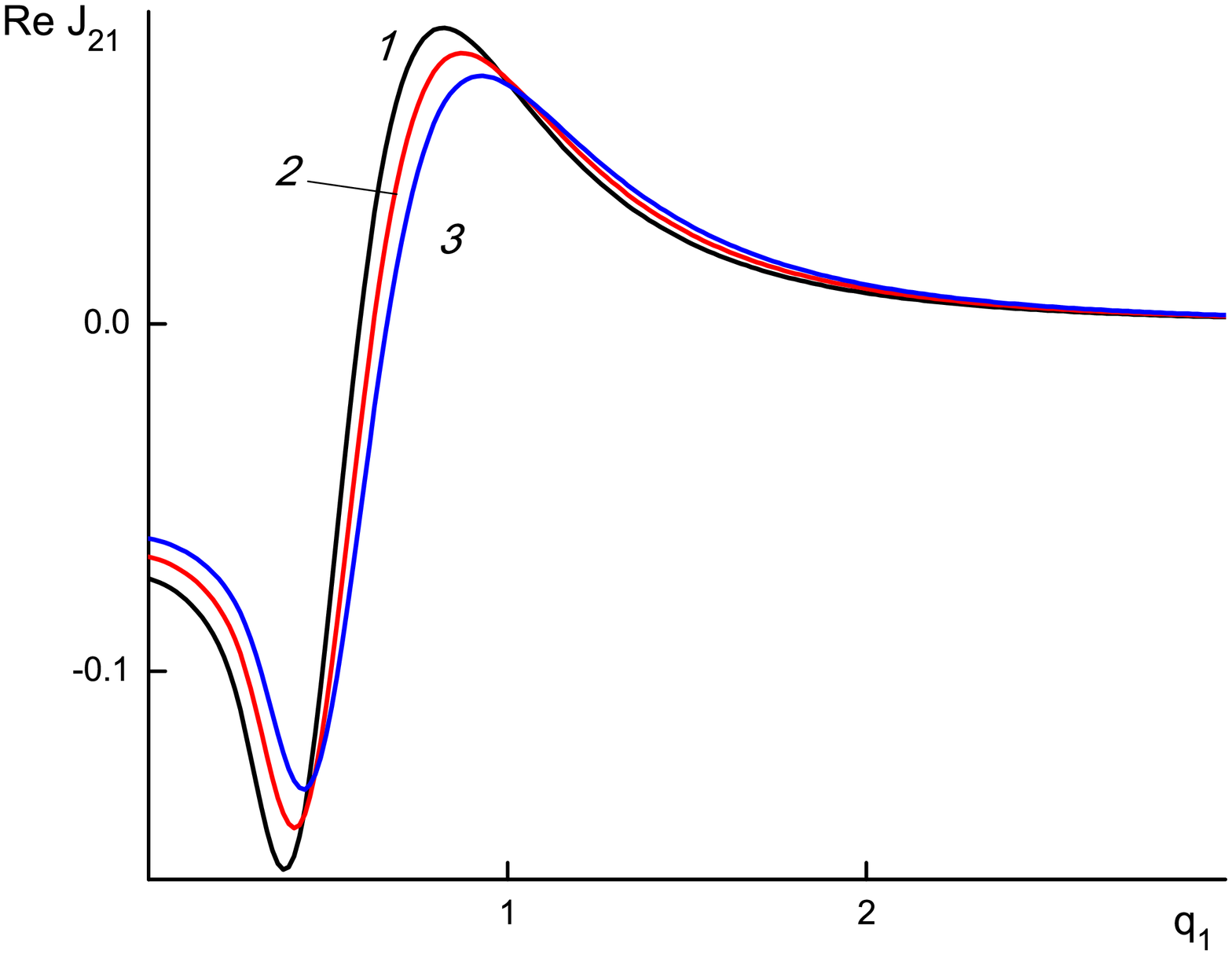}
\center{Fig. 7. Real  part of dimensionless density of longitudinal
"crossed"\, current $J_{21}$, $\Omega_1=1, q_2=0.1$.
Curves $1,2,3$ correspond to values of
dimensionless oscillation frequency of second electromagnetic field
$\Omega_2=0.1, 0.2, 0.3$.}
\includegraphics[width=17.0cm, height=10cm]{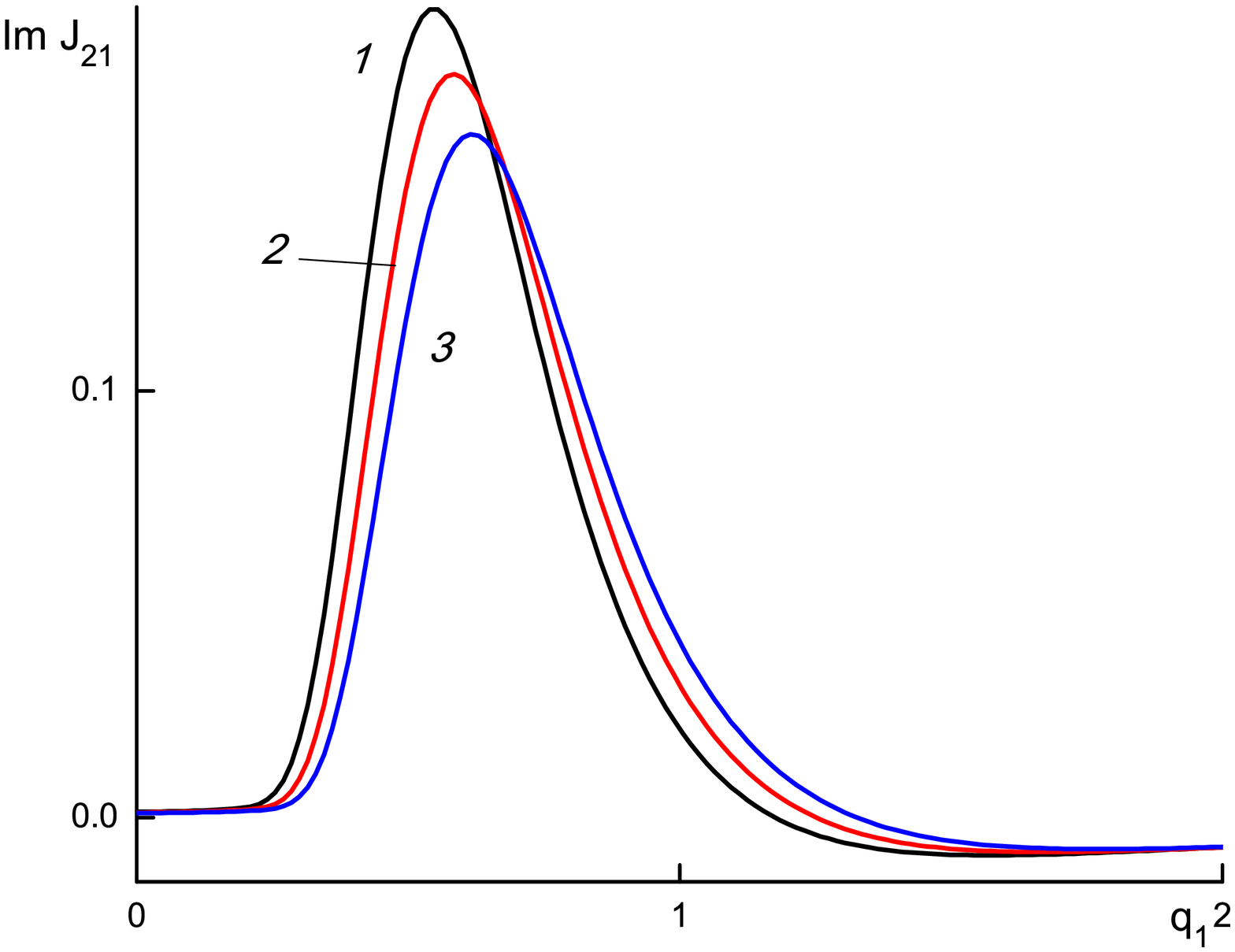}
\center{Fig. 8. Imaginary  part of dimensionless density of longitudinal
"crossed"\, current $J_{21}$, $\Omega_1=1, q_2=0.1$.
Curves $1,2,3$ correspond to values of
dimensionless oscillation frequency of second electromagnetic field
$\Omega_2=0.1, 0.2, 0.3$.}
\end{figure}


\clearpage

\begin{thebibliography}{99}
\renewcommand{\baselinestretch}{0.1}

\bibitem{BGK}{\it Bhatnagar P.L., Gross E.M., Krook M.}
Model for collision processes in gases.
I. Small amplitude processes in charged and neutral
one component systems// Phys. Rev. 1954. V. 94. P. 511--525.

\bibitem{Welander}{\it Welander P.} On the temperature jump in
rarefied gas// Arkiv for Fysik. 1954. Bd. 7. № 44. P. 507--564.

\bibitem{Gins}
{\it Ginsburg V.L., Gurevich A.V.}
The nonlinear phenomena in the plasma which is
in the variable electromagnetic
field//Uspekhy Fiz. Nauk, {\bf 70}(2) 1960; p. 201-246 (in
Russian).

\bibitem{Zyt2}{\it Kovrizhkhykh L.M. and Tsytovich V.N.}
Effects of transverse electromagnetic wave decay in a
plasma//Soviet physics JETP. 1965. V. 20. \No 4, 978-983.

\bibitem{Zyt}{\it Zytovich V.N.} Nonlinear effects in plasmas//
Uspekhy Fiz. Nauk, {\bf 90}(3) 1966; p. 435-489 (in Russian).

\bibitem{Zyt3} {\it Zytovich V.N.} Nonlinear effects in plasmas.
Moscow. Publ. Leland. 2014. 287 p. (in Russian).

\bibitem{Shukla1}{\it Shukla P. K. and Eliasson B.}
Nonlinear aspects of quantum plasma physics //
Uspekhy Fiz. Nauk, {\bf 53}(1) 2010;
[V. 180. No. 1, 55-82 (2010) (in Russian)].

\bibitem{Shukla2} {\it Eliasson B. and Shukla P. K.}
Dispersion properties of
electrostatic oscillations in quantum plasmas //
arXiv:0911.4594v1 [physics.plasm-ph] 24 Nov 2009, 9 pp.



\bibitem{Lat1}{\it Latyshev A.V. and Yushkanov A.A.}
Transverse Electric Conductivity in Collisional Quantum Plasma//
Plas\-ma Physics Report, 2012, Vol. 38, No. 11, pp. 899--908.

\bibitem{Andres}{\it De Andr\'{e}s P., Monreal R., and Flores F.}
{ Relaxation--time effects in the transverse
dielectric function and the
electromagnetic properties of metallic surfaces and small particles} //
Phys. Rev. {\bf B}. 1986. Vol. 34,\No 10, 7365--7366.


\bibitem{Gelder} {\it Gelder van, A.P.}{Quantum Corrections in the
Theory of the Anomalous Skin Effect} //
Phys. Rev. 1969. Vol. 187. \No 3. P. 833--842.

\bibitem{Fuchs}{\it Fuchs R. and Kliewer K.L.} Surface plasmon in a
semi--infinite free--electron gas //
Phys. Rev. B. 1971. V. 3. \No 7. P. 2270--2278.

\bibitem{Brod}{\it Brodin G., Marklund M., Manfredi G.}
{Quantum Plasma Effects in the Classical Regime} //
Phys. Rev. Letters. {\bf 100}, (2008). P. 175001-1 -- 175001-4.

\bibitem{Manf} {\it Manfredi G.} How to model quantum plasmas//
Topics in Kinetic Theory (Toronto, Canada, 24-26 March 2004),
Fields Inst. Comm. {\bf 46}, eds. T. Passot, C. Sulem,
P.L. Sulem (Eds), 2005, 263-287; arXiv: quant-ph/0505004.


\bibitem{Mermin} {\it Mermin N. D.}
{ Lindhard Dielectric Functions in the Relaxation--Time Approximation}.
Phys. Rev. B. 1970. V. 1, No. 5. P. 2362--2363.

\bibitem{Lat2}{\it Latyshev A. V. and Yushkanov A. A.}
Transverse electrical conductivity of a quantum collisional
plasma in the Mermin approach // Theor. and Math. Phys., {\bf
175}(1): 559--569 (2013).

\bibitem{Lat3}{\it Latyshev A. V. and Yushkanov A. A.}
Longitudinal Dielectric Permeability of a
Quntum Degenerate Plasma with a Constant Collision Frequency//
High Temperature, 2014, Vol. 52, \No 1, pp.
128--128.

\bibitem{Lat4}{\it Latyshev A. V. and Yushkanov A. A.}
Longitudinal electric conductivity in a quantum
plasma with a variable collision frequency in the framework of
the Mermin approach// Theor. and Mathem. Physics, {\bf 178}(1):
131-142 (2014).

\bibitem{Lat5}{\it Latyshev A. V. and Yushkanov A. A.}
Transverse Permittivity of Quantum Collisional
Plasma with an Arbitrary Collision Frequency//ISSN 1063-780X,
Plasma Physics Reports, 2014, Vol. 40, No. 7, pp. 564-571.


\bibitem{Lat6}{\it Latyshev A. V. and Yushkanov A. A.}
Nonlinear phenomena of generation of longitudinal electric
current by transversal electromagnetic field in plasmas//
arXiv:1502.04581v1 [phy\-si\-cs.plasm-ph], 16 Feb 2015, 16 p.


\bibitem{Lat7}{\it Latyshev A. V. and Yushkanov A. A.}
Generation of the longitudinal current by the
transversal electromagnetic field in classical and quantum
plasmas//arXiv: 1503.02102 [physics.plasm-ph] 6 Mar 2015, 27 p.

\bibitem{Lat8}{\it Latyshev A. V. and Yushkanov A. A.}
Generation of longitudinal
electric current by transversal electromagnetic field in
Maxwellian plasmas// arXiv: 1503.04478 [physics.plasm-ph] 15 Mar 2015, 18  p.

\bibitem{Lat9}
{\it Latyshev A. V. and Yushkanov A. A.}
Nonlinear longitudinal current in degenerate plasma,
arising under the influence of the transversal electromagnetic field
// arXiv:1504.05650v1 [physics.plasm-ph] 22 Apr 2015, 21  p.





\end{thebibliography}
\end{document}